%% file: main.tex
\documentclass[conference]{tex/IEEEtran}
\input{sections/packages}

\input{sections/macros}

\input{sections/highlight_solidity}
\input{sections/data}

\author{
	\IEEEauthorblockN{
		Priyanka Bose,
		Dipanjan Das,
		Fabio Gritti,
		Nicola Ruaro,
		Christopher Kruegel, and
		Giovanni Vigna
	}

	\IEEEauthorblockA{University of California, Santa Barbara}
	\IEEEauthorblockA{\{priyanka, dipanjan, degrigis, ruaronicola, chris, vigna\}@cs.ucsb.edu}
}

\begin{document}
	\title{Exploiting Unfair Advantages:
    Investigating Opportunistic Trading in the NFT Market}
	\maketitle
	\input{sections/abstract}
	\input{sections/introduction}

	\input{sections/background}
	\input{sections/analysis_approach}

	\input{sections/acquire}

	\input{sections/profit_generation}

	\input{sections/loss_minimization}
	\input{sections/related_work}
	\input{sections/conclusion}
	\bibliographystyle{plain}

\input{main.bbl}
	\clearpage
	\appendices
	\input{sections/appendix_addresses}
	\input{sections/appendix_transactions}
\end{document}

%% file: sections/packages.tex
\usepackage[switch,columnwise]{lineno}
\usepackage[linesnumbered,ruled,vlined]{algorithm2e}
\usepackage[noend]{algpseudocode}
\usepackage{amsmath}
\usepackage{stmaryrd}
\usepackage{url}
\usepackage{listings}
\usepackage{mdwlist}
\usepackage{pifont}
\usepackage{graphicx}
\usepackage{wrapfig}
\usepackage{fancyvrb}
\usepackage{paralist}
\usepackage{caption}
\usepackage{comment}
\usepackage{bbm}
\usepackage{dsfont}
\usepackage{mathrsfs}
\usepackage[T1]{fontenc}
\usepackage[dvipsnames]{xcolor}
\usepackage{esvect}
\usepackage{array}
\usepackage{multirow}
\usepackage{rotating}
\usepackage{makecell}
\usepackage{amssymb}
\usepackage{hyperref}
\usepackage{amsfonts}
\usepackage{xparse}
\usepackage[export]{adjustbox}
\usepackage{tabularx}
\usepackage{acronym}

\usepackage[]{soul}
\usepackage[]{todonotes}
\usepackage{float}
\usepackage[framemethod=tikz]{mdframed}
\usepackage{mathpartir}
\usepackage{amsthm}
\usepackage{mfirstuc}
\usepackage{booktabs}
\usepackage{colortbl}
\usepackage{stackengine}
\usepackage{xspace}
\usepackage{afterpage}
\usepackage[nomessages]{fp}
\usepackage{xfp,xintfrac}
\usepackage[autolanguage]{numprint}
\usepackage{fmtcount}
\usepackage{breqn}
\usepackage[subtle]{tex/savetrees}
\usepackage{microtype}

\mdfdefinestyle{graybox}{
	roundcorner=1mm,%
	backgroundcolor=gray!30
}

\usepackage[outline]{contour}

\graphicspath{{figures/}}

\tikzset{inlinenotestyle/.append style={align=justify}}
\SetNoFillComment
\DontPrintSemicolon

\SetCommentSty{mycommfont}
\SetFuncSty{texttt}

\newcolumntype{C}[1]{>{\centering\arraybackslash}m{#1}}
\newcolumntype{Y}{>{\raggedleft\arraybackslash}X}
\newcolumntype{Z}{>{\centering\arraybackslash}X}

\SetAlgoSkip{}

\lstdefinestyle{code}{
	caption=,
	basicstyle=\ttfamily\scriptsize,       %
	numbers=left,                   %
	numberstyle=\tiny\color{gray},  %
	stepnumber=1,                   %
	numbersep=5pt,                  %
	showspaces=false,               %
	showstringspaces=false,         %
	showtabs=false,                 %
	rulecolor=\color{black},        %
	tabsize=4,                      %
	captionpos=b,                   %
	breaklines=true,                %
	breakatwhitespace=false,        %
	morekeywords={open,read,write,close}, 		%
	keywordstyle=\bfseries\color{blue},         %
	commentstyle=\itshape\color{BrickRed},      %
	identifierstyle=\color{teal},
	stringstyle=\color{orange},         		%
	escapeinside={\%*}{*)},         %
	escapechar=|,
	belowcaptionskip=1\baselineskip,
	breaklines=true,
	frame=L,
	framesep=15pt,
	framerule=0pt,
	xleftmargin=1.5em,
	framexleftmargin=1.5em
}

%% file: sections/macros.tex
\newcommand{\etal}{\textit{et. al.}\xspace}
\newcommand{\wrt}{with respect to\xspace}
\newcommand{\etc}{\textit{etc.}\xspace}
\newcommand{\eg}{\textit{e.g.}\xspace}
\newcommand{\viz}{\textit{viz.}\xspace}
\newcommand{\ie}{\textit{i.e.}\xspace}
\newcommand{\vs}{\textit{vs.}\xspace}

\newcommand{\mypar}[1]{\vspace{1mm}\noindent\textbf{#1}\xspace}		%
\newcommand{\code}[1]{\texttt{#1}\xspace}			%

\newcommand{\Tbl}[1]{\ensuremath{\sf Table~\ref{#1}}}
\newcommand{\Fig}[1]{\ensuremath{\sf Figure~\ref{#1}}}
\newcommand{\Sec}[1]{\ensuremath{\sf Section~\ref{#1}}}
\newcommand{\Appen}[1]{\ensuremath{\sf Appendix~\ref{#1}}}

\NewDocumentCommand{\lstp}{m  m O{}}{\ensuremath{\sf Listings~\ref{#1} ~\text{and} ~\ref{#2}}}
\NewDocumentCommand{\Lines}{m  m O{}}{\ensuremath{\sf Lines~ #1\text{--}#2}}

\newcommand{\defnote}[3]{
	\newcounter{#1}
	
	\expandafter
	\newcommand\csname cmt#1\endcsname[1]{
		\refstepcounter{#1} 
		\textcolor{#3}{\textbf{#2 [\csname the#1\endcsname]:}} %
		\hl{\textbf{##1.}}} %
	
	\expandafter
	\newcommand\csname cmtdel#1\endcsname[2]{
		\refstepcounter{#1} 
		\textcolor{#3}{\textbf{#2 [\csname the#1\endcsname]:} ##2} %
		\st{\textbf{##1}}} %

	\expandafter
	\newcommand\csname #1\endcsname[1]{%
		\refstepcounter{#1}%
		{%
			\todo[color={#3!30},inline]{%
				\textbf{#2 [\csname the#1\endcsname]:}~##1}%
	}}
}

\defnote{dd}{Dipanjan}{amber}
\defnote{pb}{Priyanka}{blue}
\defnote{gv}{Giovanni}{aqua}
\defnote{ck}{Chris}{asparagus}

\let\num\numprint
\npthousandsep{,}

\definecolor{bgcode}{gray}{0.95}
\definecolor{darkorchid}{rgb}{0.6, 0.2, 0.8}
\definecolor{amber}{rgb}{1.0, 0.75, 0.0}
\definecolor{apricot}{rgb}{0.98, 0.81, 0.69}
\definecolor{aqua}{rgb}{0.0, 1.0, 1.0}
\definecolor{asparagus}{rgb}{0.53, 0.66, 0.42}
\definecolor{boxgreen}{HTML}{38761d}
\definecolor{deepgreen}{HTML}{259D04}
\definecolor{boxred}{HTML}{ff0000}
\definecolor{instrblue}{HTML}{cfe2f3}
\definecolor{instrgreen}{HTML}{b6d7a8}
\definecolor{edgered}{HTML}{ff0000}
\definecolor{hgreen}{RGB}{160,255,218}
\definecolor{hred}{RGB}{255,204,204}
\definecolor{babypink}{rgb}{0.96, 0.76, 0.76}
\definecolor{arylideyellow}{rgb}{0.91, 0.84, 0.42}
\definecolor{caribbeangreen}{rgb}{0.0, 0.8, 0.6}
\definecolor{emerald}{rgb}{0.31, 0.78, 0.47}
\definecolor{bulgarianrose}{rgb}{0.59, 0.0, 0.09}

\makeatletter
\def\endthebibliography{%
	\def\@noitemerr{\@latex@warning{Empty `thebibliography' environment}}%
	\endlist
}
\makeatother

%% file: sections/highlight_solidity.tex
\definecolor{verylightgray}{rgb}{.99,.99,.99}
\definecolor{burgundy}{rgb}{0.5, 0.0, 0.13}
\definecolor{brown(web)}{rgb}{0.65, 0.16, 0.16}

\lstdefinelanguage{Solidity}{
	keywords=[1]{anonymous, assembly, assert, balance, break, call, callcode, case, catch, class, constant, continue, constructor, contract, debugger, default, delegatecall, delete, do, else, emit, event, experimental, export, external, false, finally, for, function, gas, if, implements, import, in, indexed, instanceof, interface, internal, is, length, library, log0, log1, log2, log3, log4, memory, modifier, new, payable, pragma, private, protected, public, pure, push, require, return, returns, revert, selfdestruct, send, solidity, storage, struct, suicide, super, switch, then, this, throw, transfer, true, try, typeof, using, value, view, while, with, addmod, ecrecover, keccak256, mulmod, ripemd160, sha256, sha3, , mapping}, %
	keywordstyle=[1]\color{blue},
	keywords=[2]{address, bool, byte, bytes, bytes1, bytes2, bytes3, bytes4, bytes5, bytes6, bytes7, bytes8, bytes9, bytes10, bytes11, bytes12, bytes13, bytes14, bytes15, bytes16, bytes17, bytes18, bytes19, bytes20, bytes21, bytes22, bytes23, bytes24, bytes25, bytes26, bytes27, bytes28, bytes29, bytes30, bytes31, bytes32, enum, int, int8, int16, int24, int32, int40, int48, int56, int64, int72, int80, int88, int96, int104, int112, int120, int128, int136, int144, int152, int160, int168, int176, int184, int192, int200, int208, int216, int224, int232, int240, int248, int256, string, uint, uint8, uint16, uint24, uint32, uint40, uint48, uint56, uint64, uint72, uint80, uint88, uint96, uint104, uint112, uint120, uint128, uint136, uint144, uint152, uint160, uint168, uint176, uint184, uint192, uint200, uint208, uint216, uint224, uint232, uint240, uint248, uint256, var, void, ether, finney, szabo, wei, days, hours, minutes, seconds, weeks, years},	%
	keywordstyle=[2]\color{teal},
	keywords=[3]{block, blockhash, coinbase, difficulty, gaslimit, number, timestamp, msg, data, gas, sig, value, now, tx, gasprice, origin},	%
	keywordstyle=[3]\color{violet} ,
	identifierstyle=\color{black},
	sensitive=false,
	comment=[l]{//},
	morecomment=[s]{/*}{*/},
	commentstyle=\color{brown(web)}\ttfamily,
	stringstyle=\color{red}\ttfamily,
	morestring=[b]',
	morestring=[b]"
}

%% file: sections/data.tex
\newcommand{\bc}{blockchain\xspace}
\newcommand{\tx}{transaction}
\newcommand{\fr}{frontrunning\xspace}
\newcommand{\br}{backrunning\xspace}
\newcommand{\smart}{smart contract}
\newcommand{\solidity}{Solidity\xspace}

\newcommand{\ether}{Ether}
\newcommand{\ethereum}{Ethereum\xspace}
\newcommand{\nftm}{NFTM}
\newcommand{\ercfungible}{ERC-20\xspace}
\newcommand{\ercnonfungible}{ERC-721\xspace}
\newcommand{\ercnonfungiblemulti}{ERC-1155\xspace}
\newcommand{\dapp}{dApp}
\newcommand{\defi}{DeFi\xspace}
\newcommand{\dex}{DEX\xspace}
\newcommand{\cc}{{cryptocurrency}\xspace}
\newcommand{\opt}{{opportunistic trading}\xspace}
\newcommand{\trac}{trade action}

\newcommand{\opensea}{{\sc OpenSea}\xspace}
\newcommand{\looksrare}{{\sc LooksRare}\xspace}
\newcommand{\xy}{{\sc X2Y2}\xspace}
\newcommand{\nfttwenty}{{\sc NFT20}\xspace}
\newcommand{\nftx}{{\sc NFTX}\xspace}
\newcommand{\sorare}{{\sc Sorare}\xspace}

\newcommand{\decentraland}{{\sc Decentraland}\xspace}

\newcommand{\superrare}{{\sc SuperRare}\xspace}
\newcommand{\nifty}{{\sc Nifty}\xspace}
\newcommand{\etherscan}{{\sc Etherscan}\xspace}

\newcommand{\uniswap}{{\sc Uniswap}\xspace}
\newcommand{\sushiswap}{{\sc Sushiswap}\xspace}
\newcommand{\vtoken}{{\sc vToken}}
\newcommand{\curve}{{\sc Curve}\xspace}
\newcommand{\balancer}{{\sc Balancer}\xspace}
\newcommand{\dappradar}{{\sc DappRadar}\xspace}
\newcommand{\coingecko}{{\sc CoinGecko}\xspace}

\newcommand{\weth}{{\sc WETH}\xspace}

\newcommand{\cryptokitties}{{\sc CryptoKitties}\xspace}
\newcommand{\cryptopunks}{{\sc CryptoPunks}\xspace}
\newcommand{\tubbycats}{{\sc Tubby Cats}\xspace}
\newcommand{\azuki}{{\sc Azuki}\xspace}
\newcommand{\shibaspace}{{\sc ShibaSpace}\xspace}
\newcommand{\spaceshibas}{{\sc SpaceShibas}\xspace}
\newcommand{\clonex}{{\sc CloneX}\xspace}
\newcommand{\adidas}{{\sc Adidas Originals}\xspace}
\newcommand{\pranksy}{{\sc pranksy.eth}\xspace}
\newcommand{\mayc}{{\sc MAYC}\xspace}
\newcommand{\maycland}{{\sc MAYC Land}\xspace}
\newcommand{\aave}{{\sc Aave}\xspace}
\newcommand{\moonbirds}{{\sc Moonbirds}\xspace}
\newcommand{\mhclub}{{\sc Minimalist Human Club}\xspace}

\newcommand{\twitter}{{\sc Twitter}\xspace}

\newcommand{\flashbots}{{\sc Flashbots}\xspace}

\newcommand{\numNftMarketplaces}{5}
\newcommand{\manuallyAnalyzedNftSales}{100}
\newcommand{\nftSalevalueThresholdForManualAnalysis}{5}
\newcommand{\analysisEndBlock}{17050000}
\newcommand{\analysisStartDate}{July 15, 2022}
\newcommand{\analysisEndDate}{April 15, 2023}

\newcommand{\allTimeTradingVolumeOpenSea}{$34.69$B}
\newcommand{\allTimeTradingVolumeLooksRare}{$4.84$B}
\newcommand{\allTimeTradingVolumeCryptoPunks}{$3.1$B}
\newcommand{\allTimeTradingVolumeXY}{$1.24$B}
\newcommand{\allTimeTradingVolumeCryptoKitties}{$30.1$M}

\newcommand{\actionListing}{$\mathsf{listing}$\xspace}
\newcommand{\actionCancelListing}{$\mathsf{cancel\_listing}$\xspace}
\newcommand{\actionBuy}{$\mathsf{buy}$\xspace}

\newcommand{\numErcNFContracts}{152478}
\newcommand{\numErcMultiContracts}{27824}
\newcommand{\identifiedNftContractSampledForManualVerification}{50}
\newcommand{\identifiedNftContractTP}{50}

\FPeval{\pctIdentifiedNftContractTP}{round(\identifiedNftContractTP/\identifiedNftContractSampledForManualVerification*100, 2)}

\FPeval{\pctErcNFContracts}{round(\numErcNFContracts / (\numErcNFContracts + \numErcMultiContracts) * 100, 2)}

\newcommand{\speculationLookBackPeriod}{three\xspace}	%
\newcommand{\speculationTradeThreshold}{50}

\newcommand{\frontrunningBuyBuyOpenSea}{274123}
\newcommand{\frontrunningBuyBuyLooksRare}{1459}
\newcommand{\frontrunningBuyBuyCryptoPunks}{53}
\newcommand{\frontrunningBuyBuyXY}{40623}
\newcommand{\frontrunningBuyBuyCryptoKitties}{3649}

\newcommand{\frontrunningBuyCancelOpenSea}{4640}
\newcommand{\frontrunningBuyCancelLooksRare}{3}
\newcommand{\frontrunningBuyCancelCryptoPunks}{2}
\newcommand{\frontrunningBuyCancelXY}{1}
\newcommand{\frontrunningBuyCancelCryptoKitties}{33}

\newcommand{\frontrunningAcceptBidCancelBidOpenSea}{167}
\newcommand{\frontrunningAcceptBidCancelBidLooksRare}{0}
\newcommand{\frontrunningAcceptBidCancelBidCryptoPunks}{0}
\newcommand{\frontrunningAcceptBidCancelBidXY}{0}
\newcommand{\frontrunningAcceptBidCancelBidCryptoKitties}{0}

\newcommand{\frontrunningPlaceBidAcceptBidOpenSea}{0}
\newcommand{\frontrunningPlaceBidAcceptBidLooksRare}{0}
\newcommand{\frontrunningPlaceBidAcceptBidCryptoPunks}{200}
\newcommand{\frontrunningPlaceBidAcceptBidXY}{0}
\newcommand{\frontrunningPlaceBidAcceptBidCryptoKitties}{0}

\newcommand{\uniqueFrontrunner}{189269}

\newcommand{\attemptsPerFrontrunInstanceBinOne}{276268}
\newcommand{\attemptsPerFrontrunInstanceBinTwo}{37748}

\newcommand{\attemptsPerFrontrunInstanceBinFive}{737}
\newcommand{\attemptsPerFrontrunInstanceBinSix}{204}
\newcommand{\attemptsPerFrontrunInstanceBinSeven}{56}

\newcommand{\frontrunnersInvolvedInFrontrunningBinSix}{2186}
\newcommand{\frontrunnersInvolvedInFrontrunningBinSeven}{966}
\newcommand{\frontrunnersInvolvedInFrontrunningBinEight}{354}

\newcommand{\nonFlashbotFrontrunning}{952783}
\newcommand{\flashbotFrontrunning}{13694}
\newcommand{\pvtMiningFrontrunning}{58}
\newcommand{\nonFlashbotFrontrunningSuccess}{312972}
\newcommand{\flashbotFrontrunningSuccess}{11923}
\newcommand{\pvtMiningFrontrunningSuccess}{58}
\newcommand{\feesNonFlashbotFrontrunning}{13826.41}
\newcommand{\feesFlashbotFrontrunning}{1563.96}
\newcommand{\feesPvtMiningFrontrunning}{10}

\newcommand{\nftAcquiredThroughFrontrunningSold}{206497}
\newcommand{\nftAcquiredThroughFrontrunningSoldMadeProfit}{26513.35}	%
\newcommand{\nftAcquiredThroughFrontrunningSoldInLoss}{75078}
\newcommand{\nftAcquiredThroughFrontrunningSoldIncurredLoss}{17518.57}	%
\newcommand{\pctHoldTimeLessThanAMonth}{88}
\newcommand{\pctHoldTimeLessThanAMonthOfThoseMadeProft}{93.47}

\newcommand{\nftAcquiredThroughFrontrunningCouldNotSpeculativelyPrice}{91088}
\newcommand{\nftAcquiredThroughFrontrunningCouldSpeculativelyPrice}{19451}
\newcommand{\nftAcquiredThroughFrontrunningCouldSpeculativelyPriceNegativeProfit}{17810}
\newcommand{\nftAcquiredThroughFrontrunningCouldNotSpeculativelyPriceTotalBuyValue}{22226.84}	%

\FPeval{\nftAcquiredThroughFrontrunningSoldInProfit}{clip(\nftAcquiredThroughFrontrunningSold - \nftAcquiredThroughFrontrunningSoldInLoss)}

\FPeval{\pctNftAcquiredThroughFrontrunningSoldInProfit}{round(\nftAcquiredThroughFrontrunningSoldInProfit / \nftAcquiredThroughFrontrunningSold * 100, 2)}

\FPeval{\pctNftAcquiredThroughFrontrunningSoldInLoss}{round(\nftAcquiredThroughFrontrunningSoldInLoss / \nftAcquiredThroughFrontrunningSold * 100, 2)}

\FPeval{\totalFrontrunningBuyBuy}{clip(\frontrunningBuyBuyOpenSea + \frontrunningBuyBuyLooksRare + \frontrunningBuyBuyCryptoPunks + \frontrunningBuyBuyXY + \frontrunningBuyBuyCryptoKitties)}

\FPeval{\totalFrontrunningBuyCancel}{clip(\frontrunningBuyCancelOpenSea + \frontrunningBuyCancelLooksRare + \frontrunningBuyCancelCryptoPunks + \frontrunningBuyCancelXY + \frontrunningBuyCancelCryptoKitties)}

\FPeval{\totalFrontrunningPlaceBidAcceptBid}{clip(\frontrunningPlaceBidAcceptBidOpenSea + \frontrunningPlaceBidAcceptBidLooksRare + \frontrunningPlaceBidAcceptBidCryptoPunks + \frontrunningPlaceBidAcceptBidXY + \frontrunningPlaceBidAcceptBidCryptoKitties)}

\FPeval{\totalFrontrunningAcceptBidCancelBid}{clip(\frontrunningAcceptBidCancelBidOpenSea + \frontrunningAcceptBidCancelBidLooksRare + \frontrunningAcceptBidCancelBidCryptoPunks + \frontrunningAcceptBidCancelBidXY + \frontrunningAcceptBidCancelBidCryptoKitties)}

\FPeval{\totalFrontrunning}{clip(\totalFrontrunningBuyBuy + \totalFrontrunningBuyCancel + \totalFrontrunningAcceptBidCancelBid + \totalFrontrunningPlaceBidAcceptBid)}

\FPeval{\nftAcquiredThroughFrontrunningNotSold}{clip(\totalFrontrunning - \nftAcquiredThroughFrontrunningSold)}

\FPeval{\pctNftAcquiredThroughFrontrunningCouldNotSpeculativelyPrice}{round(\nftAcquiredThroughFrontrunningCouldNotSpeculativelyPrice / \nftAcquiredThroughFrontrunningNotSold * 100, 2)}

\FPeval{\nftAcquiredThroughFrontrunningSpeculativePricingError}{clip(\nftAcquiredThroughFrontrunningNotSold - \nftAcquiredThroughFrontrunningCouldSpeculativelyPrice - \nftAcquiredThroughFrontrunningCouldNotSpeculativelyPrice)}

\FPeval{\pctNftAcquiredThroughFrontrunningCouldSpeculativelyPriceNegativeProfit}{round(\nftAcquiredThroughFrontrunningCouldSpeculativelyPriceNegativeProfit / \nftAcquiredThroughFrontrunningCouldSpeculativelyPrice * 100, 2)}

\FPeval{\pctFrontrunningBuyBuy}{round(\totalFrontrunningBuyBuy / \totalFrontrunning * 100, 2)}

\FPeval{\pctAttemptsPerFrontrunInstanceLtSeven}{round((\attemptsPerFrontrunInstanceBinOne + \attemptsPerFrontrunInstanceBinTwo) / \totalFrontrunning * 100, 2)}

\FPeval{\attemptsPerFrontrunInstanceGtThirtyTwo}{clip(\attemptsPerFrontrunInstanceBinFive + \attemptsPerFrontrunInstanceBinSix + \attemptsPerFrontrunInstanceBinSeven)}

\FPeval{\frontrunnersInvolvedInFrontrunningGtThirtyTwo}{clip(\frontrunnersInvolvedInFrontrunningBinSix + \frontrunnersInvolvedInFrontrunningBinSeven + \frontrunnersInvolvedInFrontrunningBinEight)}

\FPeval{\pctFlashbotsFrontrunningSuccess}{round(\flashbotFrontrunningSuccess / \flashbotFrontrunning * 100, 2)}

\FPeval{\pctNonFlashbotsFrontrunningSuccess}{round(\nonFlashbotFrontrunningSuccess / \nonFlashbotFrontrunning * 100, 2)}

\FPeval{\feesPerFlashbotsFrontrunning}{round(\feesFlashbotFrontrunning / \flashbotFrontrunning, 2)}

\FPeval{\feesPerNonFlashbotsFrontrunning}{round(\feesNonFlashbotFrontrunning / \nonFlashbotFrontrunning, 2)}

\FPeval{\timesFlashbotsFrontrunningPaysThanNonFlashbots}{round(\feesPerFlashbotsFrontrunning/ \feesPerNonFlashbotsFrontrunning, 0)}

\FPeval{\pctNftAcquiredThroughFrontrunningSold}{round(\nftAcquiredThroughFrontrunningSold / \totalFrontrunning * 100, 2)}

\newcommand{\backrunningMarketplaces}{2}

\newcommand{\backrunningListingBuyCryptoPunks}{130}
\newcommand{\backrunningCryptoPunksBuyPriceLessThanAveragePrice}{94}
\newcommand{\backrunningCryptoPunksBuyPriceOneEthMoreThanAveragePrice}{21}

\newcommand{\backrunningListingBuyCryptoKitties}{181}
\newcommand{\backrunningCryptoKittiesBuyPriceLessThanAveragePrice}{99}

\FPeval{\totalBackrunningListingBuy}{clip(\backrunningListingBuyCryptoPunks + \backrunningListingBuyCryptoKitties)}

\newcommand{\mintingMethods}{86698}
\newcommand{\publicMintingMethods}{25504}
\newcommand{\mintingLimitFoundForMethods}{22857}
\newcommand{\mintingLimitViolations}{4650}

\newcommand{\mintingSampledMethods}{15}
\newcommand{\mintingUnprivilegedMethodsCouldRecoverLimitManually}{14}
\newcommand{\mintingUnprivilegedMethodsCouldRecoverLimitCorrectly}{9}

\newcommand{\corneringThresholdPercentage}{10}	%
\newcommand{\corneringThresholdTotalSupply}{50}
\newcommand{\corneredCollectionsIncludingCreators}{34221}
\newcommand{\corneredCollections}{17775}
\newcommand{\corneringHolders}{49354}
\newcommand{\corneredNft}{78904}
\newcommand{\corneredNftBought}{72881}
\newcommand{\corneredNftSold}{17967}
\newcommand{\corneringHoldersHoldingCorneringAtLeastTenCollections}{348}
\newcommand{\coreneringHolderCorneringMaximumCollections}{232}
\newcommand{\sampledAccountsOfTopHolders}{10}
\newcommand{\sampledCollectionsOfTopHolders}{5}

\newcommand{\corneringSampledReports}{10}

\FPeval{\pctCorneredNftSold}{round(\corneredNftSold / \corneredNft * 100, 2)}

\newcommand{\thresholdUniqueSenderToDetectExchange}{200}
\newcommand{\arbitrages}{31926}
\newcommand{\rewardTokenCollection}{407}
\newcommand{\profitFromaArbitrage}{4634.09}					%
\newcommand{\arbitrageProfitMinimum}{0.000005}				%
\newcommand{\arbitrageProfitMaximum}{179.2}					%
\newcommand{\arbitrageProfitTwentyFivePercentile}{0.011}	%
\newcommand{\arbitrageProfitFiftyPercentile}{0.025}			%
\newcommand{\arbitrageProfitSeventyFivePercentile}{0.069}	%

\newcommand{\arbitrageSampledReports}{20}

\newcommand{\frontrunningCancelBuyOpenSea}{3502}
\newcommand{\frontrunningCancelBuyLooksRare}{83}
\newcommand{\frontrunningCancelBuyCryptoPunks}{8}
\newcommand{\frontrunningCancelBuyXY}{356}
\newcommand{\frontrunningCancelBuyCryptoKitties}{25}

\FPeval{\totalFrontrunningCancelBuy}{clip(\frontrunningCancelBuyOpenSea + \frontrunningCancelBuyLooksRare + \frontrunningCancelBuyCryptoPunks + \frontrunningCancelBuyXY + \frontrunningCancelBuyCryptoKitties)}

%% file: sections/abstract.tex
\begin{abstract}
Since the inception of cryptocurrencies, trading has been a major use case.
As \cc evolved, new financial instruments, such as lending and borrowing protocols, currency exchanges, fungible and non-fungible tokens (NFT), staking and mining protocols have emerged.
A financial ecosystem built on top of a \bc is supposed to be fair and transparent for each participating actor.
Yet, there are sophisticated actors who turn their domain knowledge and market inefficiencies to their strategic advantage; thus extracting value from trades not accessible to others. 
This situation is further exacerbated by the fact that blockchain-based markets and decentralized finance (DeFi) instruments are mostly unregulated.
Though a large body of work has already studied the unfairness of different aspects of DeFi and \cc trading, the economic intricacies of non-fungible token (NFT) trades necessitate further analysis and academic scrutiny.

The trading volume of NFTs has skyrocketed in recent years. %
A single NFT trade worth over a million US dollars, or marketplaces making billions in revenue is not uncommon nowadays.
While previous research indicated the presence of wrongdoings in the NFT market, to our knowledge, we are the first to study predatory trading practices, what we call \textit{\opt}, in depth.
Opportunistic traders are sophisticated actors who employ automated, high-frequency NFT trading strategies, which, oftentimes, are malicious, deceptive, or, at the very least, unfair.
Such attackers weaponize their advanced technical knowledge and superior understanding of \defi protocols to disrupt trades of unsuspecting users, and collect profits from economic situations that are inaccessible to ordinary users, in a ``supposedly'' fair market.
In this paper, we explore three such broad classes of \textit{opportunistic} strategies aiming to realize three distinct trading objectives, \viz, acquire, instant profit generation, and loss minimization.
\end{abstract}

%% file: sections/introduction.tex
\section{Introduction}

Decentralized Finance (\defi), the financial system built around {\bc}s, eliminates intermediaries and provides a trustless environment to its participants.
The fact that {\bc}s are transparent means that their users have access to both code (\smart s) and data (transactions).
The promise of \textit{fairness} is built into the \bc, by design.
The trustless and transparent nature of a \bc should not only reduce the risk of fraud and manipulation, but also create a financial environment accessible equally to all the involved parties.

Unlike traditional financial markets, \defi largely lacks comprehensive regulatory frameworks, partly because it is challenging for regulatory authorities to establish consistent guidelines, and enforce them across jurisdictions.
The lack of regulation combined with the public availability of information have enabled sophisticated players with advanced technical expertise to spot high-value trade opportunities, and grab them in no time; all before the rest of the market can react.
For example, in a sandwich attack, an attacker first observes a large buy/sell order placed by a victim (transparency of the order book) that could cause significant price movements for an asset.
Then, the attacker places their own buy/sell orders on both sides of the original order.
In effect, they not only make their trade execute at a more favorable price, but also make the victim trade execute at a worse than expected price, effectively making profit out of the victim's loss.
Since the steps involved are time-sensitive and compute-intensive, such attacks can only be run by \textit{bots} (automated programs).
Evidently, such opportunities are not available to ordinary users.

Existing research on cryptoeconomics that focuses on fairness of the \cc market, economic risks, and abuses, primarily extends in two directions: 
\textbf{(i)} high-frequency trading of \ercfungible tokens, \eg, arbitrage~\cite{McLaughlin23,Zhou21,Wang22}, miner/block extractable value (MEV/BEV)~\cite{Qin21,Piet22,Bartoletti22}, frontrunning~\cite{Daian19,Eskandari20,Zhou21a2mm}, flashloan~\cite{Qin20}, 
sandwich~\cite{Zhou20} attacks, \etc, and
\textbf{(ii)} manipulation of both \ercfungible and \ercnonfungible token markets, \eg, NFT rug pull~\cite{Saharoy23,Huang23,Sharma23}, ponzi schemes~\cite{Bartoletti20,Chen19,Chen21,Kell21}, \cc pump and dump~\cite{Kamps18,Gandal18,Xu2019}, NFT trading malpractices~\cite{Das22,Vonwachter22,Lamorgia23,Bonifazi23,Serneels23,Liu23,Wen23}, \etc{}
However, none of the previous research studied the landscape of \opt in the NFT (\ercnonfungible) market.
Our paper fills that void.

Since \ercfungible tokens predate NFTs in terms of inception and popularity, therefore there exists a large body of work on analyzing predatory trading practices in that market.
However, even though NFTs are high value assets, such economic attacks in the context of NFT trading has not received enough attention yet.
For example, somebody made \$100K through speculative trading of \cryptokitties, and roughly \$8K by running an arbitrage bot~\cite{cryptokitties-arbitrage}.
On a separate occasion, a sniper bot launched a flashloan attack by \fr a \cryptopunks bid transaction, and collected a punk worth $26.25$ ETH at a mere $1$ Wei ($1\;\text{ETH} = 10^{18} \;\text{Wei}$) ~\cite{cryptopunks-frontrun} as the reward.
Such deceptive strategies not only deprive users of the profit they expected from legitimated trades, but also instill fear and mistrust within the user community, discouraging their active participation in the ecosystem.

Analyzing NFT trades or capitalizing on them come with their unique challenges for both the trader and the analyst---%
\textbf{(i) Determining profitability.}
NFTs are non-fungible tokens, and hence, no two NFTs are the same.
This makes it hard to determine the ``true'' price of a particular token, which is heavily influenced by market sentiments.
For a trader, knowing the worth of an NFT is crucial to make economically sound decisions, \eg, if it is worth to buy an NFT, if it is better to sell it off immediately, or hold it for a longer time period expecting a price increase, \etc{}
Likewise, when it comes to assessing the prices for NFTs, an analyst, too, faces similar challenges in absence of any standardized, token-specific price oracle.
\textbf{(ii) Diverse trade actions.}
Unlike \ercfungible tokens, which only support limited actions such as buy/sell, NFT trade actions are more diverse, for example, they include listing, auctions (place/accept/cancel bids), \etc{}
These actions are often marketplace-specific.
Therefore, it is hard to build an infrastructure that supports all the existing and newer marketplaces for either real-time detection, or post-mortem analysis.
\textbf{(iii) Spotting trade opportunity.}
Typically, a \ercfungible trade opportunity spans across multiple but similar types of protocols.
For example, an arbitrage can be spotted by detecting cycles in token exchange rates across decentralized exchanges~\cite{McLaughlin23}.
As we will see, profit-making trades involving NFTs typically touch multiple different types of protocols.
Detecting such opportunities in real-time is difficult in a competitive market.
As a side effect, a supposedly open (and fair) market turns out to be profitable only to advanced traders.
Likewise, the intricate and multi-protocol nature of these trades poses a challenge for analysts when it comes to identifying those opportunities.

In this paper, we study the high-frequency, \opt strategies in the NFT market.
We first categorize the opportunistic trades into three classes, then build models that identify instances of such trades, and quantify their prevalence and financial impact.
By shedding light on the under-explored or unexplored aspects of NFT trades, we make the first step to draw the attention of the community to these predatory trading practices.
In particular, we make the following contributions:

\mypar{Taxonomy of opportunistic NFT trading strategies.}
We study previous instances of opportunistic trades, and categorized them according to the strategies employed by the traders, \viz, acquire, instant profit generation, and loss minimization (\Sec{sec:approach}).
Such strategies are often malicious, deceptive, and disrupt trades of the legitimate users, while the attackers make profit out of the loss incurred by the unsuspecting users.

\mypar{Analysis of acquire strategies.}
We analyze strategies used to acquire high-value NFTs for long-term holding.
NFTs acquired this way are meant to be held for long, and sold when the market is assessed to be favorable (\Sec{sec:acquire}).

\mypar{Analysis of instant profit generation strategies.}
We analyze strategies used to make an instant profit in the NFT market.
NFTs acquired this way are meant to be held for a short time, and are typically sold in the same transaction, thus making an instant profit (\Sec{sec:instant_profit}).

\mypar{Analysis of loss minimization strategies.}
We analyze strategies used to minimize potential losses incurred from an NFT purchase  (\Sec{sec:loss_minimization}).

\mypar{Releasing code and data.}
We will release our code and data to further future research.

%% file: sections/background.tex
\section{Background}
\label{sec:background}

In this section, we will discuss concepts to understand the rest of this paper.

\mypar{Blockchain and smart contracts.}
As a distributed digital ledger, the \bc provides the necessary underpinning for \cc, a form of ownable digital asset.
Transactions change the state of the \bc.
Any connected node in the network can broadcast \tx s.
Since the participating nodes are geographically distributed, they receive those \tx s at different times and in different orders.
Therefore, a consensus protocol is necessary to achieve an agreement about the order in which transactions are executed and the ledger's state.
The native currency of a \bc, \eg, \ether{} in case of the \ethereum{} \bc, is intertwined with the functioning of the \bc itself, such as sending \tx s, offering incentives, and governance.
In addition to storing records, some blockchains can run Turing-complete programs, called \smart s.

\mypar{\emakefirstuc{\ethereum} \bc.}
\ethereum{} is one of the most popular blockchains in the world.
The current version (v$2.0$) of \ethereum follows a consensus mechanism called \textit{proof of stake} (PoS).
An \ethereum \textit{account} is an entity represented by an address that can send \tx s.
An externally owned account (EOA) is managed by a user who holds the private key associated with their address and who can initiate transactions, while contract accounts are managed by \smart s.
In PoS, \tx s are validated by special nodes called \textit{validators}, who stake (lock in) Ethers in the network to participate in the validation process.
To incentivize the validators and prevent denial-of-service (DoS) attacks, \tx s cost \textit{gas}, which is deducted from the sender's \ether{} balance.

\ethereum \smart s are written in languages like \solidity, and run on top of the \ethereum Virtual Machine (EVM).
When a contract executes, it can emit \textit{events} to indicate the occurrence of certain actions.
When \code{amount} worth of funds are sent to an address \code{to}, a contract can emit a \code{fundSent} event: \lstinline[language=Solidity]{event fundSent(address indexed to, uint amount)}.
Events are stored in \tx{} logs.
An event parameter can be marked as \textit{indexed} to allow for efficient searching and filtering.
In our example, \code{to} is indexed, while \textit{amount} is not.
Events in the \tx{} log of a contract are associated with a monotonically increasing identifier called \textit{log index}, which is assigned in the order the events are emitted.

\mypar{Decentralized finance (\defi).}
The \smart s play two important roles in the decentralized economic framework built on top of \ethereum: 
\textbf{(a)} they create and manage digital assets, called \textit{tokens}.
For example, Tether (USDT) is a fungible \ercfungible token, while \decentraland is a non-fungible \ercnonfungible token.
In the same way as the primary (native) currency, these secondary assets are also equally ownable and tradable.
\textbf{(b)} They implement decentralized protocols to automate financial processes, \eg, lending and borrowing protocols, decentralized exchanges (\dex), \etc{}
They also enable flash loan, which is a type of uncollateralized loan that instantly gives access to large amount of funds to the borrower, without needing to pay any upfront collateral.
The loan has to be settled within the same \tx{}, or else the entire \tx{} reverts.
This open financial system built around the \ethereum \bc is often called Decentralized Finance (\defi).

\mypar{Non-fungible token (NFT).}
All copies of a fungible token are identical, while every non-fungible token (NFT) is unique.
Among the NFTs managed by a \smart, each NFT is uniquely identified by a \code{tokenId}.
The creation and destruction of a token is referred to as \textit{minting} and \textit{burning}, respectively.
\ercnonfungible and \ercnonfungiblemulti are two popular standards for NFTs.
While \ercnonfungible only allows the creation of  unique tokens, \ercnonfungiblemulti is a multi-token standard that supports both fungible and non-fungible tokens within the same contract.
In effect, an NFT (identified by a \code{tokenId}) can have only one copy in a \ercnonfungible contract, while it can have multiple copies in \ercnonfungiblemulti.
The \code{totalSupply} variable in a contract keeps track of the number of unique NFTs in circulation.
A \textit{collection} refers to a set of NFTs that are managed by one \smart, and that typically have something in common, for example, they may share certain attributes, or be used for the same purpose.
NFTs in one collection typically look alike, such as \cryptopunks.
\textit{External} collections are the ones created by \smart s that are created outside of the \nftm s.

\mypar{NFT marketplace.}
NFT marketplaces (NFTMs), such as \opensea, \superrare, and \sorare connect sellers to buyers.
A seller either lists (T1) their items (NFTs) for sale, or puts them on auction (T2).
A buyers in turn, can either buy (T3) an item at the listed price, or make offers (T4), or place bids (T5) if the item is on auction.
Should they change their mind, a buyer (bidder) can also retract (T6) their bids if they have not been filled yet.
Sometimes, instead of making an offer on an individual NFT, a buyer can make a \textit{collection offer} (T7) for all NFTs in a collection.
Collection offers are useful if a buyer would like to buy any NFT in a collection, but does not have a specific NFT in mind.
A given collection offer can only be accepted once, and will expire for all other NFTs in the collection after being accepted.
We call T1-T7 \textit{trade actions}.

\mypar{NFT liquidity pool.}
The NFTMs discussed above follow a traditional orderbook-based model, where the platform maintains the list of all open orders.
Unlike orderbook-based exchanges that directly connect buyers to sellers, automated market markers (AMMs), also known as decentralized exchanges (DEX), facilitate near-instant token swaps by algorithmically setting prices for assets based on supply and demand.
Funds are sold to and bought from liquidity pools (LP), which are \smart s that lock up large volume of crowdsourced tokens.
Examples of popular \ercfungible DEXes are \uniswap, \curve, \balancer, \etc{}
Similarly, an NFT AMM is a platform that allows traders to instantly buy or sell their NFTs through LPs.
While traditional NFTMs allow the seller to have better control on the price of the item to be sold, NFT LPs are better suited for cases where the seller wants to make an instant trade.

Let's take \nftx~\cite{nftx} as an example, which is a constant-product AMM.
On \nftx, NFT holders deposit their NFTs into a vault, and mint a fungible token (\vtoken) specific to that NFT collection, for example, \textsc{PUNK} for \cryptopunks, in return.
That \vtoken represents a 1:1 claim on any one NFT from that collection's vault.
If and when the user wants to claim any NFT from the vault, they can return their \vtoken and redeem an NFT from the collection (in the pool).
Note that the \vtoken{} represents a claim on any one NFT from the specific collection, not the exact one the holder initially deposited.

LPs consist of a token pair, where one token can be exchanged for the other.
\vtoken s can be deposited into an AMM such as \sushiswap to create a liquid market.
Consider a \vtoken-ETH pool with $10$ \vtoken s and $2$ ETH.
This would make the price of the corresponding NFT $2/10 = 0.20$ ETH.
Now, if one NFT is purchased for $0.20$ ETH, that NFT is removed from one side of the pool, while $0.20$ ETH get added onto the other side, resulting in $9$ NFTs and $2.2$ ETH.
With this, the price of the NFT increases to $2.2/9 = 0.24$ ETH.

%% file: sections/analysis_approach.tex
\section{Analysis approach}
\label{sec:approach}

This paper studies \opt in the NFT market.
Traditional trading strategies tend to be based on fundamental analysis, such as studying a company's financial statements or economic indicators, and making investment decisions based on the expected long-term value of an asset.
On the other hand, \opt relies on (potentially very short-term) trends and patterns in the market.
It differs from traditional trading strategies in that it mostly seeks to take advantage of short-term market inefficiencies, news events, hype, volatility, temporary imbalances in supply and demand, price discrepancies, \etc, rather than attempting to profit from long-term trends.
While traditional trading is proactive, where the trader strives to anticipate long-term market conditions, \opt is far more reactive, where a trader is forced to react quickly to instantaneous market situations.

In this paper, we attempt to answer the following research questions: 
\textbf{(RQ1)} What types of \opt instances are prevalent in the NFT market?
\textbf{(RQ2)} How prevalent are such trades?
\textbf{(RQ3)} How profitable are these trades?

\mypar{Attacker model.}
In our model, an attacker is a trader (buyer or seller) in the NFT market with solid financial and technical knowledge, and have access to the capital and computing resources required to perform high-frequency, bot-driven trades.
The strategies they employ exhibit one or more of the following characteristics:
\textbf{(i) Malicious.}
Negatively impacts the broader ecosystem, for example, manipulating the price of tokens, market cornering, and so on.
\textbf{(ii) Disruptive.}
Their trades interfere with specific trades from ordinary users, making the latter ones fail, such as \fr.
\textbf{(iii) Deceptive.}
They bait unsuspecting users with lucrative opportunities to engage in trades that ultimately benefit the attacker, while resulting in losses for the user, for instance, enticing the user to purchase an NFT by offering it at an incredibly low price.
\textbf{(iv) Unfair.}
Such trades are competitive, which requires the attacker to quickly (and automatically) analyze, and react to changing market conditions.
For majority of the users, it is impossible to either spot those short-term opportunities, or capitalize on them.

\mypar{Methodology and data collection.}
We perform both qualitative and quantitative analysis to understand the landscape.
Since manually analyzing the profitability of and the motivation behind all NFT trades is not feasible, it makes the problem of finding \opt challenging.
To address this issue to the best possible extent, we gathered instances of \opt both actively (finding ourselves) and passively (finding previous reports).
For the passive analysis, we collected previously reported instances from publicly available resources, such as blogs, forums, write-ups and \twitter keyword searches, \eg, \textit{\fr}, \textit{arbitraging}, \textit{nft-mev}, \etc.
Then, we manually filtered out irrelevant results.
Additionally, to discover new instances, we randomly sampled $\manuallyAnalyzedNftSales$ trades from the NFT sales over $\num{\nftSalevalueThresholdForManualAnalysis}$ ETH for each of the NFTMs we support, and manually analyzed for the sign of suspicious, opportunistic trades.
We believe that our hybrid approach provides sufficient coverage of the ecosystem.

We identify three broad categories of trades---%
\textbf{(i) Acquire.}
These strategies are used to collect high-value NFTs that are worth holding for a long time (\Sec{sec:acquire}).
\textbf{(ii) Instant profit generation.}
These strategies are used to generate instant profit through buying and selling of NFTs in the same \tx{}(\Sec{sec:instant_profit}).
\textbf{(iii) Loss minimization.}
These strategies are used to minimize losses from a potentially bad purchase (\Sec{sec:loss_minimization}).

\noindent
$\blacktriangleright$
\textbf{Marketplace selection.}
To include an NFTM in our analysis, we had to first understand the marketplace-specific protocol, and then develop protocol-specific \tx{} parsers, which is non-trivial.
Therefore, we carefully integrated the support for the most prominent NFTMs in our analysis.
In line with the previous work~\cite{Kai21,Das22}, we used \dappradar~\cite{dappradar}, a popular tracker for \dapp s, to select the relevant marketplaces.
Due to the popularity of \ethereum, we selected the top four (as on \analysisStartDate{}) \ethereum-backed NFTMs  ranked by all-time transaction volume---\opensea (\allTimeTradingVolumeOpenSea), \looksrare (\allTimeTradingVolumeLooksRare), \cryptopunks (\allTimeTradingVolumeCryptoPunks), and \xy (\allTimeTradingVolumeXY).
Further, we support all three versions of \opensea marketplace, \viz, v1.0, v2.0, and v3.0.
In addition, during our passive analysis, we discovered numerous reports of \opt in \cryptokitties (\allTimeTradingVolumeCryptoKitties), which is one of the older and popular collection.
We included that too in our analysis.
Note that some of our analyses, such as minting (\Sec{sec:bulkminting}) and cornering (\Sec{sec:cornering}) are marketplace-agnostic.

\noindent
$\blacktriangleright$
\textbf{Recovering \trac s.}
For the NFTMs we considered in this work, relevant NFT trade actions (\Sec{sec:background}) are visible from the \bc.
The quantitative analysis was performed on the \ethereum \bc data until \analysisEndDate{} (block $\num{\analysisEndBlock}$).

We process each block $B$ starting from the genesis block ($block\;number=0$).
Opportunistic trades can be executed either through an EOA, or if the trade involves complicated run-time logic, touches multiple protocols, and requires atomicity of a batch of \tx s, through a \textit{bot} contract.
Therefore, in order to conduct a comprehensive analysis, we consider both external and internal \tx s, which we denote as $T$.

\nftm s vary in their trading functionalities, and may even have different implementations for the same \trac s.
Therefore, we need to consider marketplace-specific implementation to infer NFT trade details, such as action types, targeted tokens, their prices, \etc, during our analysis.
\Fig{fig:trade_actions} illustrates the predicates we extract from a \tx{} $T$.
For each $T \in B$, we extract two types of information:
\textbf{(i)} generic \tx{} details, such as block number $b$, \tx{} hash $h$, \tx{} status $st$, sender $s$ and receiver $r$, gas price $gp$, gas used $gu$, \tx{} index $ind$ within the containing block $B$, \tx{} data $d$, timestamp $ts$, \etc, and 
\textbf{(ii)} NFT trade-specific information, such as \trac s, details $n$ of the NFT involved in this trade, marketplace $m$ where the trade takes place, the buy/sell/bid price $p$ of $n$, \etc

We extract generic \tx{} details from the \bc, which is then used to infer trade-specific information in the following two steps---%
\textbf{(Step-I)} We first understand the \nftm{} protocol by examining the source code of the marketplace contract, and observing past \tx s.
Thus, we identify the public methods corresponding to the \trac s, and understand the semantics of their arguments.
\textbf{(Step-II)} In \solidity, each method (having a specific signature) is hashed to a unique four-byte-long string called \textit{method selector}.
The selector is encoded in the \tx{} data when a method is invoked.
Therefore, we match the selectors that appear in the decoded \tx{} data $d$ with the ones identified in the previous step to infer \trac s.

\begin{figure}[t]
	\[\begin{array}{rlll}
	\mathsf{t\_info}(b, h, st, s, r, & :- & \text{Transaction at block $b$} \\
	gp, gu, ind, d, ts) & & \text{with hash $h$, status $st$,} \\
	& & \text{sender $s$, receiver $r$,} \\
	& & \text{gas price $gp$,timestamp $ts$,} \\
	& & \text{index $ind$ and gas used $gu$} \\
	
	\mathsf{listing}(u, m, n, p) & :- & \text{User $u$ lists an NFT $n$ at} \\
	& & \text{\nftm{} $m$ at price $p$} \\
	
	\mathsf{cancel\_listing}(u, m, p, n) & :- & \text{User $u$ cancels the listing of an} \\
	& & \text{NFT $n$ listed at $m$ at price $p$} \\
	
	\mathsf{buy}(u, m, p, n) & :- & \text{User $u$ buys an NFT} \\
	& & \text{$n$ listed at $m$ at price $p$} \\
	
	\mathsf{place\_bid}(u, m, p, n) & :- & \text{User $u$ placed a bid of value} \\
	& & \text{$p$ on an NFT $n$ listed at $m$} \\		
	
	\mathsf{accept\_bid}(u, m, p, n) & :- & \text{User $u$ accepts a bid of value $p$} \\
	& & \text{on an NFT $n$ listed at $m$} \\
	
	\mathsf{cancel\_bid}(u, m, p, n) & :- & \text{User $u$ cancels a bid of value $p$} \\
	& & \text{on an NFT $n$ listed at $m$} \\
	\end{array}\]

        \vspace{-2mm}
	\caption{Predicates extracted from \tx s}
        \vspace{-8mm}
	\label{fig:trade_actions}
\end{figure}

In addition, for a \tx{} $T$, we define the operator
\code{TradeAction(T)}, which returns the \trac{} if $T$ represents one of the supported actions, \viz, $\mathcal{A} = \{\mathsf{listing, cancel\_listing, buy, place\_bid, accept\_bid, cancel\_bid\}}$, \textit{null} otherwise.
For a particular type of trade \tx{} T, we use the \textit{dot} ($.$) notation to refer to its attributes as defined in \Fig{fig:trade_actions}.
For example, if $T_l$ is a \actionListing \tx, $T_l.m$ denotes the marketplace where the respective NFT is listed at, and so on.
Finally, $\mathcal{T}$ denotes the set of \tx s of our interest across all the supported \nftm s $\mathcal{M}$, \ie, $\mathcal{T} = \{T \;|\; T.m \in \mathcal{M} \land \code{TradeType(T)} \in \mathcal{A}\}$.
We use the notations, predicates, and operators introduced above throughout the rest of this paper.

\noindent
$\blacktriangleright$
\textbf{Token contract identification.}
We built a dataset of \ercnonfungible and \ercnonfungiblemulti token contracts for  our analyses.
Our method of token contract identification does not require the source code of the contracts.
Since only a fraction of the contracts on the \bc have their sources available~\cite{kai}, a source code-agnostic method significantly broadens the scope of our analysis.
We took a two-step approach:
\textbf{(Step-I)} We relied on the fact that a token that conforms to \ercnonfungible or \ercnonfungiblemulti always emits a \code{Transfer} event (\Sec{sec:background}) when an NFT is transferred from one account to the other.
Such \code{Transfer} event logs can be identified from the \bc in a way similar to the method selector-based approach (used for method identification) described above.
\textbf{(Step-II)} Tokens that follow the \ercfungible standard also emit a \code{Transfer} event upon the transfer of tokens.
Thus, the \code{Transfer} logs identified in the previous step contain false positives (non-NFT tokens).
However, we observed that for \ercfungible token, the third (last) argument of the \code{Transfer} event represents the \code{value} (amount) of the tokens transferred, while for both NFT standards, it represents the \code{tokenId}.
As defined in the respective standards, this particular argument is \textit{indexed} only for the NFTs, but not for the \ercfungible tokens.
We leverage this observation to identify the \code{Transfer} logs emitted by the NFT contracts, and record the corresponding addresses.

In the end, we identified $\num{\numErcNFContracts}$ \ercnonfungible and $\num{\numErcMultiContracts}$ \ercnonfungiblemulti token contracts in total.
Further, we randomly sampled $\num{\identifiedNftContractSampledForManualVerification}$ of the identified  contracts, and verified if those are, indeed, NFTs by consulting both \opensea and \etherscan.
We found that 
all of them are true positives.

\noindent
$\blacktriangleright$
\textbf{Marketplace API.}
We use \opensea~\cite{opensea} API to fetch historical sales volumes and number of trades of NFT collections.

\noindent
$\blacktriangleright$
\textbf{Price feed.}
We fetch historical price data from \coingecko~\cite{coingecko} to convert token (\cc) prices to equivalent US Dollars at the time of the respective trades.

We provide the addresses and hashes of all important EOAs/contracts and \tx s referred to in the rest of paper in \Appen{app:eoa_contract_addresses} and \Appen{app:transactions}, respectively.

%% file: sections/acquire.tex
\section{Acquire}
\label{sec:acquire}

In a traditional market, buy-and-hold is a passive investment strategy where an investor buys an asset and holds its for an extended  period, ignoring the short-term fluctuations.
Setting the tax implications (long-term \vs short-term capital gains) aside, such strategy works well when an investor 
\textbf{(i)} has firm belief in the investment potential of the asset, or 
\textbf{(ii)} is reluctant to move with the market, and thereby, is ready to forsake profits they could make by riding out the volatility.
The NFT market is inherently volatile.
Moreover, recent research~\cite{Borri22} has shown that this volatility tends to have a negative impact on the future cumulative excess returns (returns above a benchmark or expected returns) of NFT assets.
On the other hand, popular NFTs such as \cryptopunks seem to hold their values well in the long-term despite the hype-driven nature of the NFT ecosystem.

Traders employ \textit{acquire} strategies to buy NFTs in a competitive setting, and typically hold them for the long-term rather than trying to make an instant profit by selling them off.
First, we will discuss how we determine profitability of such trades both for NFTs that could and could not be sold post-purchase.
Next, we will explore relevant strategies that we identified in the NFT market.

\input{sections/acquire_profitability}
\input{sections/acquire_frontrunning}

\input{sections/acquire_backrunning}
\input{sections/acquire_bulkminting}

\input{sections/acquire_cornering}

%% file: sections/acquire_profitability.tex
\subsection{Profitability analysis}
\label{sec:profitability_analysis}

Acquiring NFTs comes with the same risks as for any other speculative investment.
An opportunistic trader expects the price for the acquired item to go up, but does not know when/if that will happen in the future.
In practice, they can potentially sell the NFT at a profit, hold onto it for an extended period, or, in some cases, sell it at a loss due to frustration.
In this section, we show how we can measure the profitability of an NFT if the acquired asset is actually sold.
For NFTs that are still being held, we quantify the hypothetical profit that could be made, or the loss the trader would incur, if the NFT would be traded on a specific date (e.g., today).

\mypar{Generalized sale detection.}
An NFT can be sold at many different \nftm s.
Each \nftm , in turn, supports many different types of sales.
For example, we have spotted $7$ types of sale APIs in \opensea. %
To identify whether an NFT has been sold, and to obtain specific details of the sale, an analysis would need to be capable of handling all these trade variations, which is infeasible.
To circumvent this issue, we designed the following marketplace and trade-agnostic analysis.

\noindent
$\blacktriangleright$
\mypar{(Step I) Detecting sales.}
Since an NFT sale transfers the ownership from the seller $s$ (trader) to the buyer $b$, it emits a \code{Transfer} event.
If the seller $s$ has purchased the NFT with \code{tokenId} at \tx{} $T_b$, then we look for the first  external \tx{} $T_s$ that
\textbf{(i)} occurs after $T_b$, and
\textbf{(ii)} emits a log of the form \code{Transfer}$(s, b, tokenId)$.
We consider $T_s$ to be the potential sale \tx.

\noindent
$\blacktriangleright$
\mypar{(Step II) Obtaining sale details.}
Given $T_s$, our goal is to obtain the sale details without any specific knowledge about the \nftm{} protocol.
The key observation is that the NFT sales are atomic, \ie, the NFT transfer and payments to the involved parties are made (through internal \tx s) in the same external \tx{} $T_s$.
To this end, we first build a payment graph $\mathcal{G}=\{e\}$.
Specifically, if any internal \tx{} of $T_s$ transfers $v$ amount of \ether or \ercfungible tokens from a payer account $pr$ to a payee account $pe$, we add a directed edge $e = \langle pr, pe, v\rangle$ to $\mathcal{G}$.

We traverse $\mathcal{G}$ recursively in a depth-first manner, starting from the buyer node $b$.
At each iteration, when we process an edge $e = \langle pr, pe, v\rangle$, we also update two quantities: \code{payIn}: total amount of buyer's money received by the seller, and \code{payOut}: total deductions from the seller's earnings, such as marketplace fees, creator royalties, \etc{}
Specifically, we add $v$ to \code{payIn}, if payee $pe$ is the seller $s$, and to \code{payOut}, if payer $pr$ is the seller $s$.
Finally, we compute the net earning of the seller by subtracting \code{payOut} from \code{payIn}.

One can argue that we could do away with the payment graph-based approach, and calculate the earnings by summing up the incoming/outgoing payments to/from the seller.
That would be imprecise, because $s$ could receive money from sources other than $b$ in the same external \tx.
With the payment graph, we are essentially taint-tracking the flow of money from $b$ (source) to $s$ (sink), thus only considering amounts involved in the trade in question.
For cases where $T_s$ is not a true sale, but an unconditional transfer of NFT from $s$ to $b$, we treat them as the same entity.
Then, we re-apply the algorithm assuming $b$ to be the seller.

\mypar{Pricing model.}
For an NFT that has not been sold, our pricing model computes an assumed price $p$ for that NFT as of a specific data (such as today), and thereby, can determine the potential profit/loss the trader could have made/incurred. %
Our objective is not to design a precise price oracle, but rather to offer an intuitive indication of whether these (still held) NFTs retain their value.  
We follow three simple rules to determine the current market value $p$ of an item:
\textbf{(i)} We expect an NFT that holds its market value should be traded at least once within last \speculationLookBackPeriod{} months.
If not, then we treat $p=0$.
\textbf{(ii)} Even if the NFT has been traded within the last \speculationLookBackPeriod{} months, we want to ignore occasional trades, which does not reflect the true market demand for that item.
Therefore, if the NFT has been traded fewer than $\speculationTradeThreshold{}$ times, we disregard those sales, and again treat $p=0$.
\textbf{(iii)} Otherwise, we consider $p$ to be the average sale price of the collection.
\opensea periodically releases the total sale volume and the number of trades within a timeframe for a collection in \textit{batches}.
The frequency of release varies based on the trade activity of that collection.
We consider the most recent such batch to compute $p$.

\mypar{Limitations.}
Our generic sale detection approach is unable to handle batch trades in which several NFTs are sold together as a bundle.
We can, of course, compute the total earnings from selling all the NFTs.
However, due to the lack of protocol knowledge, we cannot automatically infer the sale price for each specific NFT.
This is acceptable, as most of the opportunistic trades, be it a buy or a sale, target specific NFTs, not a bundle.

%% file: sections/acquire_frontrunning.tex
\subsection{\emakefirstuc{\fr}}
\emakefirstuc{\fr} refers to the malicious practice of exploiting the knowledge of a pending victim \tx{} $T_v$, and use that knowledge to execute an attack \tx{} $T_a$ before $T_v$.
By scheduling their \tx{} before the victim's \tx{}, an attacker makes a profit that the victim was supposed to make. %

The \ethereum mempool is a globally visible, temporary, public storage area of unconfirmed \tx s that are waiting to be included in a block.
When a user sends a \tx, it enters the mempool, and awaits validation by miners.
Miners typically prioritize \tx s according to the gas price offered.
The higher the gas price, the more incentive miners have to include the \tx{} in the next block they mine.
In \fr, an attacker first observes a victim \tx{} $T_v$ from the mempool.
If it deems profitable, then they replay the same \tx, but with a higher gas price, so that the miners include the attack \tx{} $T_a$ before $T_v$.
Ironically, in context of \opt, the attacker not only banks on the victim's effort of finding a profitable trade opportunity (which is already hard in the competitive market), but also robs the victim of their profit.

With the rising popularity of NFTs as an investment vehicle, \fr has become prevalent in the NFT trading as well.
Traders frontrun NFT trades to either buy or sell NFTs depending on the trade actions (\Sec{sec:background}) performed by the trades.
We identified the following types of \fr attacks.

\mypar{Buy--Buy.}
In this attack, both the victim $V$ and the attacker $A$ are buyers.
When $A$ spots a profitable buy order \tx{} $T_v$ submitted by $V$ to purchase an NFT $N$, $A$ frontruns $T_v$ with another buy order $T_a$ targeting the same NFT.
As a result, $A$ acquires the NFT, while $T_v$ becomes invalid and eventually reverts.

It is worth noting that if $A$ knew beforehand that purchasing $N$ is a worthwhile investment, then they could probably make the same purchase without even \fr $V$, just like a regular buyer.
However, the NFT market is hype-driven, and speculative.
Only a few special-purpose utility NFTs, which represent in-game assets or tickets to access services, hold intrinsic value.
For other NFTs, their valuation is largely derived from factors such as supply and demand and people's perception.
Thus, seeing the victim $V$ attempting to purchase an NFT serves as an endorsement, or signal of  interest, which triggers $A$ to frontrun the victim.

\mypar{Buy--Cancel.}
In this attack, the victim $V$ is a seller, while the attacker $A$ is a buyer.
When $A$ observes a cancel order transaction $T_v$, where $V$ attempts to cancel a previously-listed NFT, $A$ frontruns $T_v$ with a buy order $T_a$ targeting that NFT.
As a result, $A$ acquires the NFT, while $T_v$ becomes invalid and reverts.

Typically, this type of \fr attack happens when a seller unintentionally lists an NFT at an exceptionally low price, but then realizes the mistake, and promptly attempts to cancel the listing to avoid any potential loss.
Unfortunately, this very attempt of the seller to prevent a loss attracts the attention of the attacker to that NFT. %

\mypar{Accept bid--Cancel bid.}
In this attack, the victim $V$ is a buyer, while the attacker $A$ is a seller.
When $A$ observes a cancel bid transaction $T_v$ to cancel a prior bid submitted by $V$ on an NFT, $A$ frontruns $T_v$ with an accept bid \tx{} $T_a$, thus forcing $V$ to purchase the NFT against their will.

This can occur either due to the buyer's accidental bid placement, or their subsequent realization that the deal may not be beneficial.
The seller then acts quickly to exploit the situation before the buyer's bid is canceled, ensuring they do not miss out on the possibility of a profitable outcome.

\mypar{Place bid--Accept bid.}
In this attack, both the victim $V$ and the attacker $A$ are buyers.
When $A$ observes an accept bid \tx{} $T_{ab}$ from a seller to accept a bid $B$ submitted by a potential buyer $V$, $A$ frontruns $T_{ab}$ with a place bid \tx{} $T_a$ with a slightly higher bid amount than $B$, so that the NFT is sold to them instead of the victim.

In certain cases involving popular collections, there have been instances where victim buyers place  low bids $B$, most likely to try out their luck.
Then, they patiently wait for the sellers to accept those bids.
On the other hand, the seller gets frustrated by only receiving low bids, and eventually goes ahead to accept one of these low bids.
An attacker then exploits the situation by \fr the accept bid \tx{} $T_{ab}$ by placing a slightly higher bid than the existing one.
As a consequence, the original bid is removed from consideration, and the attacker's bid becomes the most recent one.
Consequently, the seller ends up accepting the bid placed by the attacker instead of the initial bid $B$ placed by the victim buyer.
Note that this attack is only possible if the marketplace protocol does not include any unique identifier corresponding to the victim bid $B$ in $T_{ab}$.
Rather, $T_{ab}$ is bid-agnostic and just accepts the currently highest bid. %
Unlike other cases, where the victim \tx{} fails as the side-effect of the frontrun, here, it is necessary that both \tx s execute successfully for the attack to work.

\input{tables/tbl_frontrunning}

\mypar{Identifying \fr attacks.}
To quantify the prevalence of the \fr attacks presented above, we apply the following set of rules to detect past instances of successful frontruns.

Let $T_1, T_2 \in \mathcal{T}$ be two \tx s, such that $T_1.ts \leq T_2.ts$.
Then, we say that $T_1$ frontruns $T_2$, if:

\vspace{-0.5mm}
\mypar{Rule 1.}
They involve the same NFT, and the same marketplace, but come from different senders.
Formally, $T_1.n = T_2.n$, and $T_1.m = T_2.m$, and $T_1.s \neq T_2.s$.

\vspace{-0.5mm}
\mypar{Rule 2.}
The \tx s are ``competing'' with each other, so they appear in the same block.
\ie, $T_1.b = T_2.b$ and $T_1.ind < T_2.ind$.
The gas price offered for $T_1$ is higher than that of $T_2$, \ie, ${T_1}.gp > {T_2}.gp$.

\vspace{-0.5mm}
\mypar{Rule 3.}
The \tx s represent the appropriate types of trade action required by the type of frontrun attack in question.
For example, if we are looking for a Buy--Cancel frontrun, then $\code{TradeType}(T_1) = $ \actionBuy and $\code{TradeType}(T_2) = $ \actionCancelListing.

\vspace{-0.5mm}
\mypar{Rule 4.}
As discussed earlier, for all types of frontrun categories except the last one (Place bid--Accept bid), the victim \tx{} fails, while the attack \tx{} succeeds.
Therefore, we perform appropriate checks on $T_1.st$ and $T_2.st$ status values.

\begin{figure}[!t]
	\centering
	\includegraphics[width=0.8\linewidth]{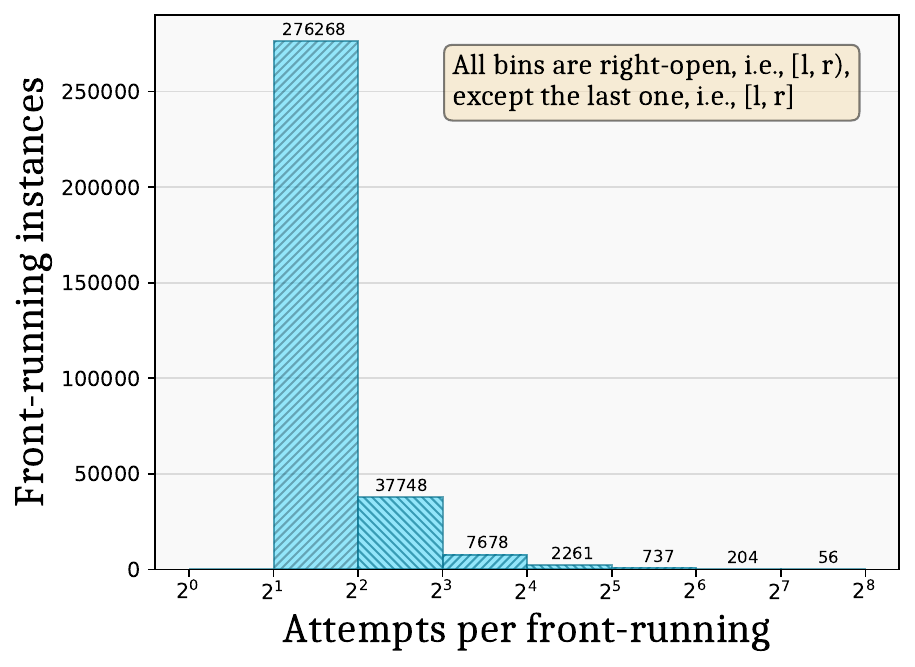}
	\vspace{-1mm}
	\caption{Distribution of number of \fr attempts per instance.
		X-axis follows a log scale}
        \vspace{-5mm}
	\label{fig:frontrun_attempts}
\end{figure}

\begin{figure}[!t]
	\centering
	\includegraphics[width=0.8\linewidth]{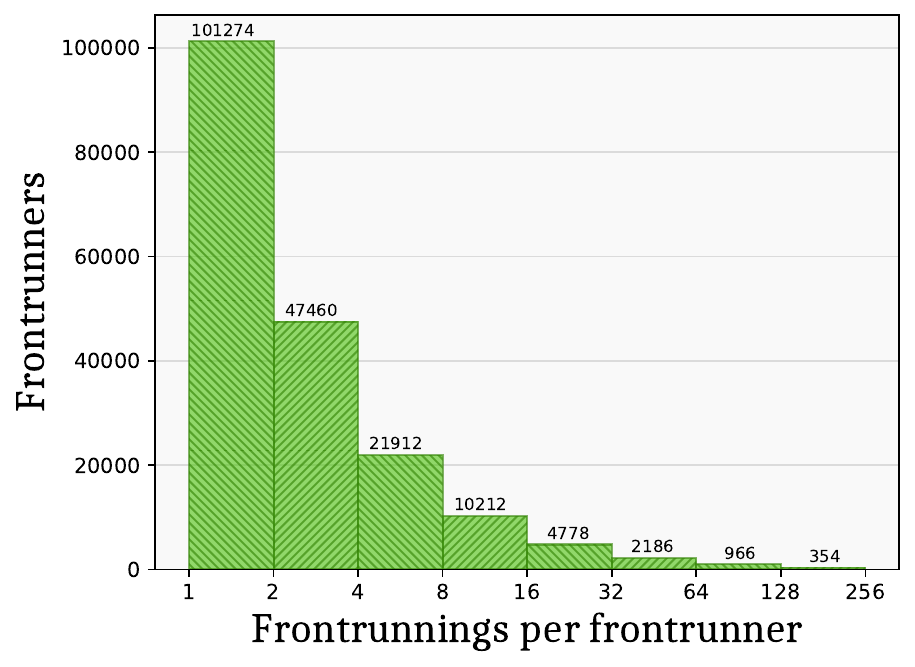}
	\caption{Distribution of \fr instances a frontrunner has participated.
		X-axis follows a log scale}
	\label{fig:frontrunner_dist}
	\vspace{-8mm}
\end{figure}

\begin{figure}[!t]
	\centering
	\includegraphics[width=0.8\linewidth]{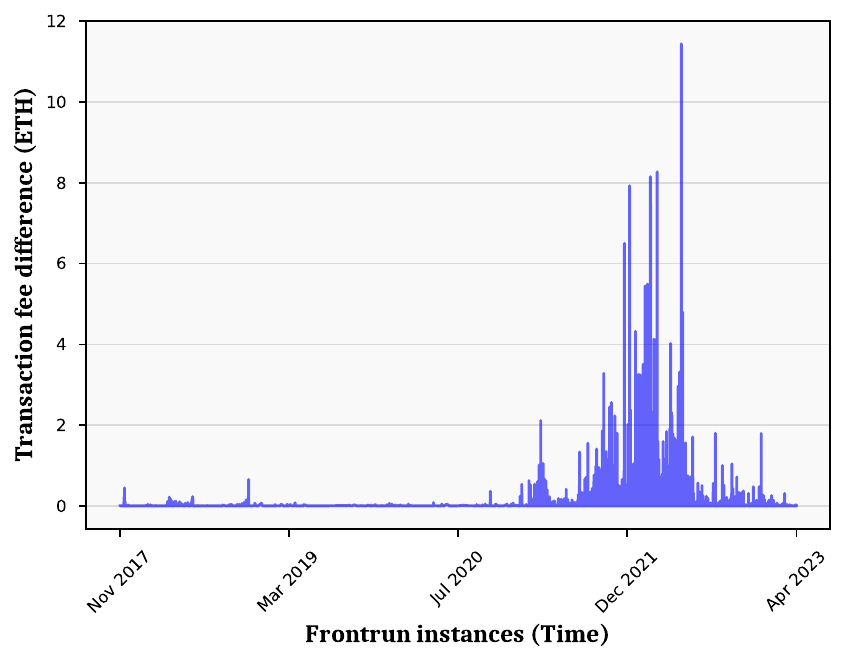}
	\caption{Difference in the gas costs of the highest (successful) and the lowest gas-paying frontrunners}
        \vspace{-7mm}
	\label{fig:gas_price_diff_on_trx_fee_dist_line}
\end{figure}

\mypar{Quantitative analysis.}
We detected $\num{\totalFrontrunning}$ instances of \fr in \numberstringnum{\numNftMarketplaces} \nftm s attempted by $\num{\uniqueFrontrunner}$ unique frontrunners.
\Tbl{tbl:frontrunning_backrunning_loss_minimization} shows the number of different \fr instances we detected per exchange.
Our results indicate that Buy-Buy is the most prevalent ($\pctFrontrunningBuyBuy\%$) type of \fr in the ecosystem.
Given the hype-driven nature of NFT market, we believe that such \textit{reactive} purchases are quite expected.

A \fr instance targets a particular NFT.
In our prior discussion, we explained frontrunning with a single attacker. 
However, it is possible that multiple frontrunners target the same NFT, in particular for very popular assets.
We plot the distribution of the number of frontrunners per attack instance in \Fig{fig:frontrun_attempts}.
The vast majority ($\pctAttemptsPerFrontrunInstanceLtSeven\%$) of the frontruns were attempted by less than $7$ frontrunners.
However, we also found  $\num{\attemptsPerFrontrunInstanceGtThirtyTwo}$ instances with more than $32$ frontrunners, implying that those trades were highly competitive.

Since a frontrunner may repeatedly engage in attacks, we asked ourselves this question: \textit{How many frontrunners out there dominate the \fr landscape?}
To this end, \Fig{fig:frontrunner_dist} shows the distribution of the number of \fr instances a frontrunner is involved in.
This figure reveals that a large number of frontrunners took part in only a few trades.
On the other hand, a small number ($\num{\frontrunnersInvolvedInFrontrunningGtThirtyTwo}$) of frontrunners are each involved in at least $32$ trades.
As one might expect, the market indeed exhibits a concentration of a small number of significant (active) players. %

As previous studies~\cite{Daian19} have pointed out, \fr may lead to reactive bidding and ``gas wars'' to become the top choice of the miners during \tx-ordering.
We were interested to see the difference in gas cost $gc_h$ incurred by successful frontrunners (who pay the highest gas price) and the gas cost ($gc_l$) that the lowest-paying frontrunners offered. %
The amount of gas used by a successful \tx{} should stay the same regardless of the sender.
Therefore, to estimate $gc_l$, we used the amount of gas used ($gu$) by a successful frontrunner, thus simulating a successful \tx{} even in the failure case.
We compute the gas costs by multiplying $gu$ with respective gas prices, and plot the difference $(gc_h - gc_l)$ in \Fig{fig:gas_price_diff_on_trx_fee_dist_line}.
As it becomes apparent, the gas war intensified between December $2021$ and August $2022$, exactly around the time when the NFT-hype took off.

To give an idea of how competitive \fr could be, this \azuki NFT was purchased through a Buy-Buy \fr in May $2022$ at \tx{} \code{0x1e8...93b}.
Surprisingly, while the frontrunner paid $0.16$ ETH for the NFT itself, they paid a whopping $12$ ETH for the gas to win the competition against $7$ other frontrunners.
Later, they were able to sell it at $14.36$ ETH on the same day.
Upon further analysis, we realized that the average price of the \azuki collection was $12$ ETH at that time.
Therefore, this was a ``steal deal'' for the buyer, which prompted them to win the gas war at a high cost.
Interestingly, the seller themselves bought that NFT at $12.5$ ETH only four hours before the listing was put up, which led us to suspect that it was an unfortunate incident of mislisting.
In another instance, while buying the very first token of the \adidas collection at \tx{} \code{0xa7f4...e59}, the buyer had to compete with $223$ other frontrunners.

\input{tables/tbl_frontrunning_miner}

Sending \tx s through the public mempool attracts unwanted attention from frontrunners.
To mitigate this risk, people resort to \flashbots-like~\cite{flashbots} \fr-protecting services that entirely bypasses the mempool.
Fortunately, \flashbots publishes a list of \tx s relayed through their service to maintain transparency.
Additionally, we detect \textit{private mining} by following any on-chain payment made directly to the block miner (bribe) by the sender in the same \tx{} as the \fr.
The number of \tx s and the fees paid to the miners in \fr attempts mined through various sources are summarized in \Tbl{tbl:frontrunning_miner}.
All privately mined \tx s were successful.
Also, \fr attempts through \flashbots have a noticeably higher success rate ($\pctFlashbotsFrontrunningSuccess$\%) than the ones mined through the mempool ($\pctNonFlashbotsFrontrunningSuccess\%$).
However, the reliability of \flashbots comes at a cost; on average, \tx s mined through \flashbots paid $\timesFlashbotsFrontrunningPaysThanNonFlashbots$ times higher fees per \tx{} than the ones mined through mempool.

An interesting question is to see how many of the NFTs acquired through \fr actually earned profits in the long run.
We found that frontrunners could sell $\num{\nftAcquiredThroughFrontrunningSold}$ ($\pctNftAcquiredThroughFrontrunningSold\%$) NFTs, of which $\num{\nftAcquiredThroughFrontrunningSoldInProfit}$ ($\pctNftAcquiredThroughFrontrunningSoldInProfit\%$) were sold for a cumulative profit of $\num{\nftAcquiredThroughFrontrunningSoldMadeProfit}$ ETH.
Sales of the remaining $\num{\nftAcquiredThroughFrontrunningSoldInLoss}$ ($\pctNftAcquiredThroughFrontrunningSoldInLoss\%$) NFTs incurred a total loss of $\num{\nftAcquiredThroughFrontrunningSoldIncurredLoss}$ ETH.
This is surprising.
Despite the widespread attention NFTs have received in recent times and the influx of traders seeking quick profits through risky trades, a notable portion of frontrunners are unable to achieve their financial goals.
For example, a \clonex NFT was purchased at \tx{} \code{0xfbb...07e} for $32$ ETH, while less than a year later, it had to be sold at $9.5$ ETH, because the collection price was rapidly declining.

Another interesting observation is that of the NFTs that were sold, $\pctHoldTimeLessThanAMonth\%$ were held less than a month.
For those that were sold at a profit,   $\pctHoldTimeLessThanAMonthOfThoseMadeProft\%$ were held less than a month.

Further, we apply our pricing model (\Sec{sec:profitability_analysis}) to all $\num{\nftAcquiredThroughFrontrunningNotSold}$ unsold NFTs.
We did not detect any sales for $\num{\nftAcquiredThroughFrontrunningCouldNotSpeculativelyPrice}$ ($\pctNftAcquiredThroughFrontrunningCouldNotSpeculativelyPrice\%$) tokens in the last \speculationLookBackPeriod months, suggesting that those tokens, with a cumulative purchase price of $\num{\nftAcquiredThroughFrontrunningCouldNotSpeculativelyPriceTotalBuyValue}$ ETH, have already gotten severely devalued since purchase.
We could speculatively price $\num{\nftAcquiredThroughFrontrunningCouldSpeculativelyPrice}$ tokens (the remaining $\num{\nftAcquiredThroughFrontrunningSpeculativePricingError}$ errored out due to minor issues), of which $\num{\nftAcquiredThroughFrontrunningCouldSpeculativelyPriceNegativeProfit}$ ($\pctNftAcquiredThroughFrontrunningCouldSpeculativelyPriceNegativeProfit\%$) already exhibit negative profit as of today.

%% file: tables/tbl_frontrunning.tex
\begin{table}[t]
	\small
	\centering
	\setlength{\tabcolsep}{0.5em}
	
	\begin{tabular}{@{}|l|r|r|r|r|r|@{}}
		\hline
		
		\multicolumn{1}{|c|}{\thead{Strategy}} & \multicolumn{5}{c|}{\thead{Marketplaces}} \\
		
		\cline{2-6}

		& \multicolumn{1}{c|}{\rotatebox{90}{\opensea}} & \multicolumn{1}{c|}{\rotatebox{90}{\looksrare}} & \multicolumn{1}{c|}{\rotatebox{90}{\cryptopunks}} & \multicolumn{1}{c|}{\rotatebox{90}{\xy}} & \multicolumn{1}{c|}{\rotatebox{90}{\cryptokitties}} \\
		
		\hline

		\rowcolor{black!10}\multicolumn{1}{|c|}{\textbf{Acquire}} & & & & & \\
		
		\textbf{\emakefirstuc{\fr}} & & & & & \\
		
		\hspace{2mm} Buy--Buy & $\num{\frontrunningBuyBuyOpenSea}$ & $\num{\frontrunningBuyBuyLooksRare}$ & $\num{\frontrunningBuyBuyCryptoPunks}$ & $\num{\frontrunningBuyBuyXY}$ & $\num{\frontrunningBuyBuyCryptoKitties}$ \\
		
		\hspace{2mm} Buy--Cancel & $\num{\frontrunningBuyCancelOpenSea}$ & $\num{\frontrunningBuyCancelLooksRare}$ & $\num{\frontrunningBuyCancelCryptoPunks}$ & $\num{\frontrunningBuyCancelXY}$ & $\num{\frontrunningBuyCancelCryptoKitties}$ \\
		
		\hspace{2mm} Accept bid--Cancel bid & $\num{\frontrunningAcceptBidCancelBidOpenSea}$ & $\num{\frontrunningAcceptBidCancelBidLooksRare}$ & $\num{\frontrunningAcceptBidCancelBidCryptoPunks}$ & $\num{\frontrunningAcceptBidCancelBidXY}$ & $\num{\frontrunningAcceptBidCancelBidCryptoKitties}$ \\ 
		
		\hspace{2mm} Place bid--Accept bid & -- &  -- & $\num{\frontrunningPlaceBidAcceptBidCryptoPunks}$ & -- & $\num{\frontrunningPlaceBidAcceptBidCryptoKitties}$ \\
		
		\textbf{\emakefirstuc{\br}} & & & & & \\
		
		\hspace{2mm} Listing--Buy & -- & -- & $\num{\backrunningListingBuyCryptoPunks}$ & -- & $\num{\backrunningListingBuyCryptoKitties}$ \\
		
		\hline
		
		\rowcolor{black!10}\multicolumn{1}{|c|}{\textbf{Loss minimization}} & & & & & \\
		
		Cancel--Buy & $\num{\frontrunningCancelBuyOpenSea}$ & $\num{\frontrunningCancelBuyLooksRare}$ & $\num{\frontrunningCancelBuyCryptoPunks}$ & $\num{\frontrunningCancelBuyXY}$ & $\num{\frontrunningCancelBuyCryptoKitties}$ \\
		
		\hline
	\end{tabular}

	\caption{Instances of \fr, \br, and loss minimizing trades found in different marketplaces}
	\label{tbl:frontrunning_backrunning_loss_minimization}
\end{table}

%% file: tables/tbl_frontrunning_miner.tex
\begin{table}[t]
	\small
	\centering
	\setlength{\tabcolsep}{0.5em}
	
	\begin{tabular}{p{2.5cm}|r|r|r}

	\hline
	
	& \thead{Non-flashbots} & \thead{Flashbots} & \thead{Private mining} \\
	
	\midrule
	
	\textbf{Frontrunnings} & $\num{\nonFlashbotFrontrunning}$ & $\num{\flashbotFrontrunning}$ & $\num{\pvtMiningFrontrunning}$ \\
	
	\textbf{\makecell[l]{Successful\\ frontrunnings}} & $\num{\nonFlashbotFrontrunningSuccess}$ & $\num{\flashbotFrontrunningSuccess}$ & $\num{\pvtMiningFrontrunningSuccess}$ \\
	
	\textbf{Miner fees (ETH)} & $\num{\feesNonFlashbotFrontrunning}$ & $\num{\feesFlashbotFrontrunning}$ & $\num{\feesPvtMiningFrontrunning}$ \\
	
	\bottomrule
	
	\end{tabular}
	
	\caption{\emakefirstuc{\fr} attempts made through \flashbots and private mining}
        \vspace{-8mm}
	\label{tbl:frontrunning_miner}
\end{table}

%% file: sections/acquire_backrunning.tex
\subsection{\emakefirstuc{\br}}

In a \br attack, the \tx{} sender $A$ gets their transaction $T_a$ to appear immediately {\it after} a target \tx{} $T_v$ sent by victim $V$.
In the same way as \fr, \br exploits the knowledge of the pending \tx s in the mempool.
In this attack, the backrunner $A$ offers a slightly lower gas price in $T_a$ than in $T_v$ so that their \tx{} gets deprioritized \wrt the target \tx{} during inclusion in the block.
We identified the following type of \opt that leverages \br to acquire an NFT.

\mypar{Listing--Buy.}
In this type of trade, the target user $V$ is a seller, and the backrunner $A$ is a buyer.
When $V$ submits a listing transaction $T_v$ for a highly desirable NFT, $A$ might spot this opportunity and immediately backrun $T_v$ with a buy \tx{} $T_a$ to acquire the NFT before others have a chance to do so.
In effect, a regular user will not be able to see the NFT listed in the marketplace, and make an informed purchase decision, since the item already got sold before it even became visible to the general public. 
For the \br to be successful, it is necessary that both the seller's and the buyer's  \tx s get executed successfully.

\begin{figure}[!t]
	\centering
	\includegraphics[width=0.8\linewidth]{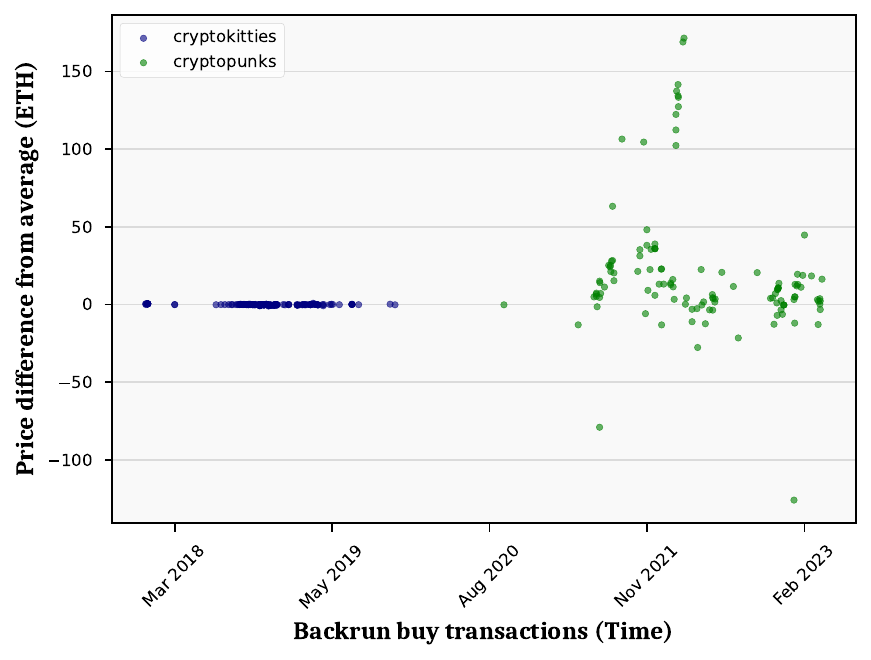}
	\caption{Difference in the average price of the backrun collection and the buy price of the NFT. A positive difference indicates that the backrun was potentially profitable}
    \label{fig:backrun_price_diff}
    \vspace{-7mm}
\end{figure}

\mypar{Quantitative analysis.}
For \br to take place, the marketplace needs to have on-chain listing so that it becomes visible in the mempool.
Since the item listings for \opensea, \looksrare, and \xy are maintained off-chain, we could only detect \br in \cryptopunks and \cryptokitties.
In those \numberstringnum{\backrunningMarketplaces} \nftm s, we detected $\num{\totalBackrunningListingBuy}$ instances of \br.

Backrunning a listing to purchase an NFT makes sense only if the NFT is bought at a reasonable price so that the buyer would be able to sell it for a profit afterwards.
In other words, paying more than the market average would hurt  profitability.
While individual buyers have their own definition of reasonable price, we used a naive oracle of the average price of that collection for further analysis.
In \Fig{fig:backrun_price_diff}, we illustrate the price difference between the average price of the  collection at the moment when  backrunning occurred, and the purchase price of the NFT.
A positive difference indicates that the item was acquired below the then-current market price, thus making the trade potentially profitable.
As we can see, most of the purchases were made close to or below the market averages.
For example, $\num{\backrunningCryptoPunksBuyPriceLessThanAveragePrice}$ \cryptopunks and $\num{\backrunningCryptoKittiesBuyPriceLessThanAveragePrice}$ \cryptokitties were bought below the market average.
In fact, among those \cryptopunks, $\num{\backrunningCryptoPunksBuyPriceOneEthMoreThanAveragePrice}$ were bought at more than one \ether{} below the market price.
This confirms the existence of \br in the context of NFT trades.

%% file: sections/acquire_bulkminting.tex
\subsection{Bulk minting}
\label{sec:bulkminting}

Minting is the process of creating new tokens.
There are primarily two ways in which NFTs are minted: 
\textbf{(i) Mint before sale (MBS).}
NFTs are minted by ``privileged'' accounts before any public sale starts.
This is done largely for three reasons:
\textbf{(i.a) Secondary sale.}
The first sale of an item by the creator is called \textit{primary sale}, while any subsequent sales by any seller other than the creator is known as \textit{secondary sale}.
For external collections, \nftm s generally act as secondary marketplaces, meaning that they do not facilitate the primary sale.
To address this issue, creators first mint the NFTs to their accounts, and then list those items on the \nftm s, like any other regular secondary sale.
\textbf{(i.b) Giveaway.}
As part of a promotion, creators often mint a few items to distribute them to selected accounts.
\textbf{(i.c) Whitelisting.}
For hyped collections, whitelisting is the way to ensure that certain early followers, patrons, contributors or enthusiasts receive NFTs, no matter how competitive the market is.
These accounts are added to a \textit{whitelist}, so that they can mint and redeem their allotted NFTs before the public minting starts.
\textbf{(ii) Mint at sale (MAS).}
In this case, the primary sale is conducted through collection-specific websites.
The minting method is unprivileged, meaning that it can be called by the public.
When a buyer purchases an NFT on that website, it is minted as part of the sale \tx and gets transferred to the buyer's account right away.
Consequently, the sale price of the NFT includes the price for the gas associated with the minting.

When a new collection is launched, a sale start date is advertised to create excitement.
In the recent past, the minting hype has been so intense that many times when a new collection was launched, it caused a surge in the gas price for all Ethereum transactions to cope with the increased volume of minting \tx s~\cite{gas-hike-during-mint-1,gas-hike-during-mint-2,gas-hike-during-mint-3}.
Opportunistic traders ride on this hype, and they might want to mint unusually large amounts of NFTs as soon as a collection is launched, depriving the rest of the market.
We call this \textit{bulk minting}.
It is particularly problematic, since it leads to non-uniform distribution of assets, which not only hurts the overall tokenomics, but may also spur distrust and disinterest among the buyers for the project.

MBS is less susceptible to bulk minting, since the minting process is carried out through privileged accounts.
Popular tokens that adopt MAS, on the other hand, often implement anti-bulk-minting measures to ensure a more fair distribution of tokens (wealth).
In \Fig{fig:anti_minting}, we show one such check present in the \tubbycats collection.
This check limits the maximum number of NFTs one can mint in a single \tx.
Unfortunately, defenses like these are not strong enough to dissuade smart traders.
On the other hand, a more stringent check, like enforcing limits on the total ownership, is undesirable, since it may hinder the growth of the project in the long run.

To evade existing anti-bulk-minting defenses, instead of minting the NFTs from an EOA, traders often leverage \smart s.
\Fig{fig:evade_anti_minting} shows one such example.
Since \tx s initiated through contracts are atomic in nature, a single invocation of \code{bulkMint()} dispatches ten calls (in the \code{for} loop) to \code{mintForSale()}, thus minting $50$ NFTs in one batch.

\begin{figure}[t]
\begin{lstlisting}[
	caption=,
	numbers=none,
	breakatwhitespace=true,
	language=Solidity]
function mintFromSale(uint tubbiesToMint) 
                                public payable {
    require(tubbiesToMint <= 5, "Only up to 5 
                tubbies can be minted at once");
    .
    .
}
\end{lstlisting}
        \vspace{-4mm}
	\caption{Anti-minting measure in \tubbycats NFT}
        \vspace{-5mm}
	\label{fig:anti_minting}
\end{figure}

\vspace{-3mm}
\begin{figure}[h]
\begin{lstlisting}[
	caption=,
	numbers=none,
	breakatwhitespace=true,
	language=Solidity]
function bulkMint() {
    require(msg.sender == 0x1337... 1337);
    for (i = 1; i < 10; i++) {
        mintFromSale(5);
    }
}
\end{lstlisting}
        \vspace{-5mm}
	\caption{Evading anti-minting measures}
        \vspace{-4mm}
	\label{fig:evade_anti_minting}
\end{figure}

This technique has two benefits: 
\textbf{(i)} since every call to \code{mintForSale()} is a separate (internal) \tx, it bypasses the mitigation presented in \Fig{fig:anti_minting}, and
\textbf{(ii)} to win in a gas war, the trader has to bid a high gas price only once to invoke \code{bulkMint()}, as opposed to incurring similar gas cost ten times if they had used an EOA to mint the tokens in ten separate transactions.

\mypar{Identifying minting limit evasions.}
To identify instances of attackers who attempt to evade minting limits, we first need to determine the minting limit for an NFT token.
We do so in three steps.

\noindent
$\blacktriangleright$
\textbf{(Step I) Minting \tx{} identification.}
When a token with \code{tokenId} is transferred from the \code{from} address to the \code{to} address, a \code{Transfer} event 
is emitted on the log: \lstinline[language=Solidity]{event Transfer(address indexed from, address indexed to, uint256 tokenId)}.
We rely on these logs to determine the time an NFT is sold, its buyer, and the number of NFTs held by the buyer at that point.
A \code{Transfer} event decrements and increments the number of tokens held by the \code{sender} (seller) as well as the \code{receiver} (buyer), respectively.
We enumerate the transfer logs emitted by an NFT contract sequentially, block by block, and within each block, ordered by log index, thus ensuring that the transfer events are considered chronologically.
At each step, we keep track of the \code{tokenIds} received by the \code{receiver}, and the number of NFTs in circulation at that point.
Therefore, when we see a \code{tokenId} that we have not seen before, we consider that as a \textit{minting} event.
We record the \tx{} that generated the log ($T_m$), add it to the set of all minting \tx s, and increase the \code{totalSupply} accordingly.

The algorithm above, in its current form, works only for \ercnonfungible tokens. \ercnonfungiblemulti is a ``token-within-tokens'' standard, because each token can have multiple copies.
Adding support for \ercnonfungiblemulti would not only  require us to extend the algorithm  by adding extra accounting for each \code{tokenId}, but it would also make the downstream analyses more complicated.
Therefore, for our minting and cornering (\Sec{sec:cornering}) analysis, we focus only on \ercnonfungible tokens.
Since it is the older of the two NFT standards, and represents $\pctErcNFContracts\%$ contracts in our dataset, we believe that it is a reasonable trade-off. 

\noindent
$\blacktriangleright$
\textbf{(Step II) Minting function identification.}
As mentioned previously, limit evasion attacks are only relevant when MAS is employed, as minting is open to general public.
We leveraged the $T{_m}s$ collected in the previous step to identify the minting methods in the NFT contract.
In particular, our goal is to identify unprivileged minting methods that can be called by any user (\ie, methods that are not restricted by owner-only checks, and/or restricted to a whitelist).
To do so, we first collect the set $F$ of all public methods that are touched by $T_m$.
To filter privileged methods included in $F$, we replay them at respective blocks on our testnet, but with a sender different from the original one, and monitor if the \tx{} reverts.
If it does, then we consider the minting function to be privileged, and discard it.
Otherwise, we add it to $F_u$, the set of all unprivileged minting methods.

\noindent
$\blacktriangleright$
\textbf{(Step III) Minting limit identification.}
Unfortunately, none of the token standards define a minting API.
However, if a minting method $f_m \in F_u$ allows minting multiple tokens in one \tx, then it should also accept an argument $a_m$ that defines the number of tokens to be minted.
Due to the absence of a standardized method signature, we first need to identify the part of the \textit{calldata} (the serialized input to $f_m$ in a \tx) that corresponds to $a_m$.
To do so, we leverage the fact that when tokens are minted, a standard-compliant contract emits \code{Transfer} logs, one for each minted token.
First, we choose a \tx{} $tx_m$ that invokes $f_m$ to mint $n$ tokens, which we determine by scanning the logs.
Then, we locate $n$ (encoded) in the calldata, thus inferring the position of $a_m$.
Finally, we replay the \tx{} with gradually increasing values of $a_m$, until the execution reverts, which we assume happens when the minting limit is reached.
We adapt binary search to converge to a solution faster.
Some contracts expect the user to send the exact amount $e$ of \ether s that is needed to mint $n$ tokens.
To tackle this issue, we estimate $e$ by computing the price per token from $tx_m$, which we then multiply with $a_m$, the number of tokens.

\mypar{Quantitative analysis.}
For the $\num{\numErcNFContracts}$ \ercnonfungible contracts, we identified $\num{\mintingMethods}$ minting methods.
Out of these, $\num{\publicMintingMethods}$ were unprivileged methods.
Our analysis could recover the minting limit for $\num{\mintingLimitFoundForMethods}$ of those methods.
Note that if a contract does not explicitly specify a limit, we default to using one.
Finally, we look for external \tx s that call those unprivileged methods and mint tokens beyond the inferred limits.
We identified $\num{\mintingLimitViolations}$ instances where minting limits were evaded.

As an example, though the \mhclub token limits minting to $2$ tokens per \tx, in \tx{} \code{0x1f8...4aa}, $200$ tokens were minted through a contract.

\mypar{Manual verification.}
We sampled $\mintingSampledMethods$ methods from the sets of methods marked as ``privileged'' and ``unprivileged'' by our analysis, respectively.
We chose contracts with source for  ease of manual verification.
All 30 methods were correctly labeled.

Next, for the $\mintingSampledMethods$ unprivileged methods, we manually went through the source code to determine the minting limits.
Due to the complexity of that logic, we could not find the limit in one contract.
For the remaining $\mintingUnprivilegedMethodsCouldRecoverLimitManually$ methods, our analysis correctly identified the limits for $\mintingUnprivilegedMethodsCouldRecoverLimitCorrectly$.
As for the false detections, we noticed that the arguments that our analysis used to bruteforce the limit were not always accurate.
Specifically, our heuristic attempts to infer the argument $a_m$ via log analysis (as explained before).
However, the signatures of some of the minting methods are far more complicated than this simple analysis can correctly handle. 
For example, certain methods accept a list of \code{tokenId}s to be minted (rather than a single one).
In general, recovering the signature of a method from contract bytecode is an open problem, tackling which goes beyond the scope of this work.

\mypar{Limitations.}
\textbf{(i)} Our heuristic to estimate $e$ is best-effort.
If the token employs dynamic pricing, then our \tx{} reverts due to a mismatch between the price and the value, thus inferring a wrong limit.
\textbf{(ii)} If the limit is influenced by dynamic factors, such as availability, then the extracted limit might be imprecise.
\textbf{(iii)} We make no effort to recover signatures of the minting methods.
Therefore, our analysis can return incorrect results for methods with complex signatures.

%% file: sections/acquire_cornering.tex
\subsection{Market cornering}
\label{sec:cornering}

Cornering is the act of obtaining enough assets of a certain type so that the holders can control the market and eventually drive up the asset's price.
Attackers often manipulate the price by buying or selling items in large quantities, creating artificial shortages of supply. %
Both the US Commodity Futures Trading Commission Securities (CFTC) and the US Securities \& Exchange Commission (SEC) consider cornering a form of illegal market manipulation under both the securities and futures laws~\cite{sec-cftc-cornering}.
However, as NFT market are unregulated, ``whales'' (a term that refers to entities that possess significant amounts of capital) can buy a large amount of tokens as soon as a collection is launched and hoard them until an opportune moment.
Since the \code{totalSupply} of a typical NFT collection is much smaller than for \ercfungible tokens, it is far easier to corner the NFT market than that of \ercfungible tokens.

\mypar{Identifying cornering instances.}
We detect instances of cornering during the minting \tx{} identification step (Step-I) in \Sec{sec:bulkminting}. 
Specifically, each time a \code{Transfer} event is processed, and a receiver (buyer) receives a token, we check the fraction $f$ of the total number of available tokens (\code{totalSupply}) this buyer owns at that point.
If $f$ is above a threshold $th_f$, it indicates that a significant portion of the tokens are under the control of that buyer, which we count as an instance of cornering.
Also, when minting for a token has just started and the \code{totalSupply} (available liquidity) is very small, the concept of cornering is not meaningful.
Hence, we do not consider any detected cornering instances until the asset's \code{totalSupply} is above a threshold $th_t$.

\mypar{Quantitative analysis.}
We empirically set $th_f = \corneringThresholdPercentage$ and $th_t = \corneringThresholdTotalSupply$ for our analysis.
That is, if we detect at any point that there are at least $\corneringThresholdTotalSupply$ tokens minted for a collection, and more than $\corneringThresholdPercentage\%$ of the total supply is being held by an address, we flag that as a cornering attempt.
Out of the $\num{\numErcNFContracts}$ \ercnonfungible contracts, we initially detected $\num{\corneredCollectionsIncludingCreators}$ collections with cornering activity.
Upon further inspection, we immediately noticed that our analysis incorrectly flagged a large number of collection creators who were minting tokens before their sale(recall the MBS model).
Thus, we excluded all cornering instances for collections with a single holder (at the time the alert was raised).
While this might remove true cornering attempts, we believe that cases where a single attacker purchases all tokens are rare. 
At the end, we are left with $\num{\corneredNft}$ cornered NFTs from $\num{\corneredCollections}$ collections cornered by $\num{\corneringHolders}$ holders.
Of the cornered NFTs, $\num{\corneredNftBought}$ were actually purchased (we could identify corresponding buy \tx).
For the remaining cases, the NFTs were unconditionally transferred, most likely from a different account (that was used for the purchase of those tokens) of the same holder.

Interestingly, a mere $\pctCorneredNftSold\%$ of the cornered tokens were later sold, or could be sold.
This led us to question: \textit{Is cornering NFTs at all profitable?}
To investigate further, we computed the distribution of the number of cornered collections per holder, and identified $\num{\corneringHoldersHoldingCorneringAtLeastTenCollections}$  accounts that are holding tokens of at least ten collections.
Notably, the account \pranksy cornered the most number of collections ($\num{\coreneringHolderCorneringMaximumCollections}$) in the past.
We randomly sampled $\sampledAccountsOfTopHolders$ accounts of top holders, and then again sampled $\sampledCollectionsOfTopHolders$ collections from each of their holdings to dig in further.
Some of those collections have already been flagged by \nftm s like \opensea, and have been taken down, leaving the holders with no market to sell off their assets.
Some other collections are still ``alive,'' but without significant movement in price since the time they were bought.
It seems that these mega-holders (whales) are buying inexpensive tokens in bulk without a meticulous market analysis.
Then, they are holding on to these tokens, hoping that some of the cornered collections will increase in price over time.

\mypar{Manual verification.}
We built two datasets for manual verification.
First, we randomly sampled $\corneringSampledReports$ holder accounts (Set-A) from the reported results.
Second, we selected the top $\corneringSampledReports$ holders (Set-B) that are cornering the most number of collections.
For the selected accounts, we checked:  
\textbf{(i)} Whether the accounts are ``special'' in any way such that holding a large amount of tokens cannot be deemed as cornering. 
\textbf{(ii)} For each collection, \etherscan shows the top token holders for that collection/tokens.
Similarly, for each address, \etherscan also shows the amount of tokens of a particular type that an address is currently holding.
However, we have no way to know the historical token holdings of an address.
Since, according to our experience, buy-and-hold is a common strategy for these accounts, we verified (consulting \etherscan) if, at least, the current holdings of an account is above our threshold.

In Set-A, we encountered the \nifty wallet, which is a false positive.
\nftm s like \nifty~\cite{nifty} follow an escrow model~\cite{Das22}, where the marketplace holds all the assets for all the users in its wallet. %
Without deeper protocol knowledge, this case is indistinguishable from a true cornering instance.
For one account, the current holding is below our threshold, which we could not further verify.
In Set-B, the top two holders were the \textit{null} address, where tokens are typically minted from, and the \textit{burn} address, which is used to destroy tokens to reduce the available liquidity.
Except those two, the remaining ones were true positives.

\mypar{Limitations.}
\nftm s that follow an escrow model would appear as attackers in our results.
However, given that such \nftm s are not common, we did not make any specific effort to exclude them in our analysis.

%% file: sections/profit_generation.tex
\section{Instant profit generation}
\label{sec:instant_profit}

Unlike \textit{acquire} strategies that come with no guarantee of future profit, instant profit generation refers to the ability to generate profits immediately from a trade.
Instead of holding onto an asset for extended amount of time and hoping that the market works in favor of the trader, with instant profit generation, a trader purchases an NFT and sells it (with profit) in the same \tx.
Because profit is acquired in the same \tx{} as the purchase, a \smart{} can often verify if a profit has indeed been made at the end of the \tx.
If no profit is detected, then the contract reverts the \tx, thus saving the trader from getting stuck with a non-profitable item.
The nature of the trade leaves no room for optimizing for later (long-term) profit, which may result in a lower profit margin.
However, it offers the advantage of assured risk-free gains, which is a significant benefit in a highly speculative market.
Next, we will explore two such strategies commonly observed in the NFT market.

\subsection{Arbitrage}

In finance, arbitrage is the process of capitalizing on the price discrepancy of an asset across different markets.
The asset is bought at a lower price on one market, and immediately sold at a higher price on another.
The profit is derived from the difference in prices.

\ercfungible coins are fungible.
All units of a coin are identical and interchangeable.
Fungibility helps in creating markets with a healthy supply and demand for the same asset.
For example, an arbitrage opportunity opens up when \textsc{MATIC} is sold and bought at slightly different prices on both \uniswap and \sushiswap, two popular decentralized exchanges. %
On the other hand, NFTs are non-fungible, which means that each unit is unique.
With traditional sale mechanisms, such as sell orders or auctions, only the NFT owner $S$ could list an item.
When a buyer $B$ purchases an item, all offers from other potential buyers, which were tied to the specific asset and its previous owner $S$, would become invalid.
Moreover, in the absence of an existing (open) buy order, a buyer $B$ has no way to sell their purchased item in the same \tx{} (and obtain the difference as profit).
Therefore, arbitrage based on NFTs used to be impossible.
This changed recently when advanced financial instruments like collection offers and liquidity pools were introduced.
Below, we describe ways in which these two mechanisms enable arbitrage in the NFT ecosystem.

\mypar{Collection offer-based arbitrages.}
\nftm s allow buyers to make offers on items to express their interest in purchasing them.
Such offers are tied to a specific NFT.
Unlike traditional offers, a \textit{collection offer} expresses an interest in \textit{any} item of a collection (\Sec{sec:background}).

Assume that a seller $S$ is selling an NFT $N$ from a collection $C$ at price $p_s$.
Also, there is a potential buyer $B$, who has made a collection offer on the same collection $C$ at price $p_b$, meaning that they are willing to purchase any NFT of $C$ at that price.
If $p_b > p_s$, an arbitrageur then buys $N$ from $S$ at $p_s$, sells it to $B$ at $p_b$ in the same \tx, and pockets the difference $(p_b - p_s)$ as the profit.

In \tx{} \code{0x0659...29d}, a \moonbirds NFT was bought from \looksrare for $44$ ETH, and sold to a collection offer for $223$ ETH on the same platform, pocketing $180$ ETH after fees.

\mypar{Liquidity pool-based arbitrages.}
NFT liquidity pools (LP) allow one to sell their NFT $N$ to a \smart{} instead of a particular buyer.
In return, the seller receives a pool-specific token $T_N$, which is tied to the NFTs of the collection where $N$ belongs to (\Sec{sec:background}).

LP-based arbitrages can be arbitrarily long and complex.
However, the basic structure is as follows:
\textbf{(Step-I)} The arbitrageur purchases an NFT $N$ from an \nftm{} at a price of $p_b$ ETH.
\textbf{(Step-II)} They sell $N$ to an LP, and receive an equivalent amount of pool tokens $T_N$.
\textbf{(Step-III)} Finally, they swap all those $T_N$ tokens at a \ercfungible \dex for $p_s$ amount of ETH.
If all the exchange rates are favorable, the arbitrageur ends up with more money than they invested, allowing them to keep the difference $(p_s - p_b)$ as profit.

In \tx{} \code{0x304...031}, the arbitrageur purchases a \spaceshibas NFT from \opensea at $0.015$ ETH.
They immediately swap it on \nfttwenty for $97$ \shibaspace \ercfungible tokens, which they then deposited into \uniswap v$2.0$ \shibaspace-\weth pool to get $0.033$ WETH back.
Even with a \tx{} fee of $0.013$ ETH, they made an instant, risk-free profit of $0.005$ ETH.

\subsection{Reward token collection}

Creators often offer reward tokens to incentivize buyers of their NFTs.
Rewards are typically announced during the initial days of the launch of a collection, to either entice users to purchase or hold  NFTs.
Reward tokens often provide additional advantages.
For instance, if a creator intends to release another collection later on, these tokens could grant holders the opportunity to acquire those NFTs at a discounted price.
For reward token collection trade, the trader sells the original NFT immediately, but keeps the reward tokens as profit.

The following steps are involved in this kind of activity:
\textbf{(Step-I)} The trader borrows $p_b$ ETH using a flash loan.
\textbf{(Step-II)} With the borrowed funds, they proceed to purchase an NFT of type $N_O$.
\textbf{(Step-III)} For purchasing $N_O$, they receive an additional NFT of type $N_R$ as the reward.
\textbf{(Step-IV)} Finally, they sell the original NFT $N_O$, either through a collection offer, or to a liquidity pool.
The proceeds from this sale are then used to repay the borrowed funds $p_b$ obtained through the flash loan.
They keep the reward token $N_R$ as the profit.

In \tx{} \code{0x629...f4d}, a \mayc token was bought for $28$ ETH using a flash loan from \aave.
The trader then redeemed a \maycland token worth $15$ ETH, and sold the original \mayc for $26.4$ ETH in a collection offer.

\mypar{Identifying instant profit generating \tx s.}
Instant profit generating trades come in different shapes.
Identifying and encoding the signature of each such trade in our analysis would be infeasible.
Instead of making specific assumptions about the ``shapes'' of such trades, we instead propose a trade-agnostic, payment flow-based approach to identify (instant) profit-generating transactions.
Let $T$ be an external \tx{} sent by a sender $S$ to a receiver $R$.

$\blacktriangleright$
\mypar{(Step I)} We consider $T$ further if it performs at least one NFT sale, and the buyer $B$ transfers the NFT out to some other account, indicating an immediate sale.

$\blacktriangleright$
\mypar{(Step II)} Let $\mathcal{A}$ be the set of addresses touched by $T$ (and its internal \tx s).
A trader perhaps controls more than one of those accounts, any of which could receive the profit from the trade.
Our goal is to build the set $\mathcal{M}$ of such addresses.
Both the addresses of the sender $S$ and buyer $B$ are included in $\mathcal{M}$.
In addition, many arbitrages are performed through \smart s (the receiver $R$), which can receive the profit, too.
Therefore, we build the initial set $\mathcal{M} = \{S, B, R\}$.

$\blacktriangleright$
\mypar{(Step III)} There could still be an address $a_t \in \mathcal{A} \land a_t \notin \mathcal{M}$, which is under the control of the trader, but not apparently visible from this \tx{} $T$.
We extend $\mathcal{M}$ to include such addresses $a_t$.
To this end, we scan through all the past \tx s of each address $a_m \in \mathcal{M}$, and add any address $a_i$ that has interacted with $a_m$ to $\mathcal{M}$.
While doing so, we exclude an $a_i$ if it is an exchange.
Specifically, we disregard any $a_i$ having more than a threshold $th_e$ (empirically determined) number of unique senders, which is common for exchanges.

$\blacktriangleright$
\mypar{(Step IV)} We only retain addresses in $\mathcal{M}$ that also appear in  \tx{} $T$, \ie, we update $\mathcal{M} = \mathcal{M} \cap \mathcal{A}$.
We consider $\mathcal{M}$ to be the set of addresses potentially controlled by the trader.
Therefore, our analysis treats them as a single ``clique,'' and disregards any mutual interactions among the addresses in $\mathcal{M}$.

$\blacktriangleright$
\mypar{(Step V)} For $\mathcal{M}$, we compute two quantities: \code{payIn}, which is the total amount of \ercfungible tokens, \ether, and NFTs flowing in, and \code{payOut}, which is the total amount of \ercfungible tokens, \ether, and NFTs flowing out.
If the net cash in-flow $(\code{payIn} - \code{payOut})$ is positive, we flag $T$ as an instant profit generating \tx.

\mypar{Quantitative analysis.} For this analysis, we used a threshold value $th_e = \thresholdUniqueSenderToDetectExchange$. 
We detected $\num{\arbitrages}$ arbitrage \tx s, making a total of $\num{\profitFromaArbitrage}$ \ether s in profit, and $\num{\rewardTokenCollection}$ instances of reward-token collection.
Among these transactions, the minimum and maximum profits were $\arbitrageProfitMinimum$ ETH and $\num{\arbitrageProfitMaximum}$ ETH, respectively.
Further, we broke down the profits by percentiles: the $25$\textsuperscript{th}, $50$\textsuperscript{th} (median), and $75$\textsuperscript{th} percentiles were  $\arbitrageProfitTwentyFivePercentile$ ETH, $\arbitrageProfitFiftyPercentile$ ETH, and $\arbitrageProfitSeventyFivePercentile$ ETH, respectively.
This suggests that most of the opportunities yield small profits, while larger, more lucrative ones are rarer.

\begin{figure}[!t]
	\centering
	\includegraphics[width=0.8\linewidth]{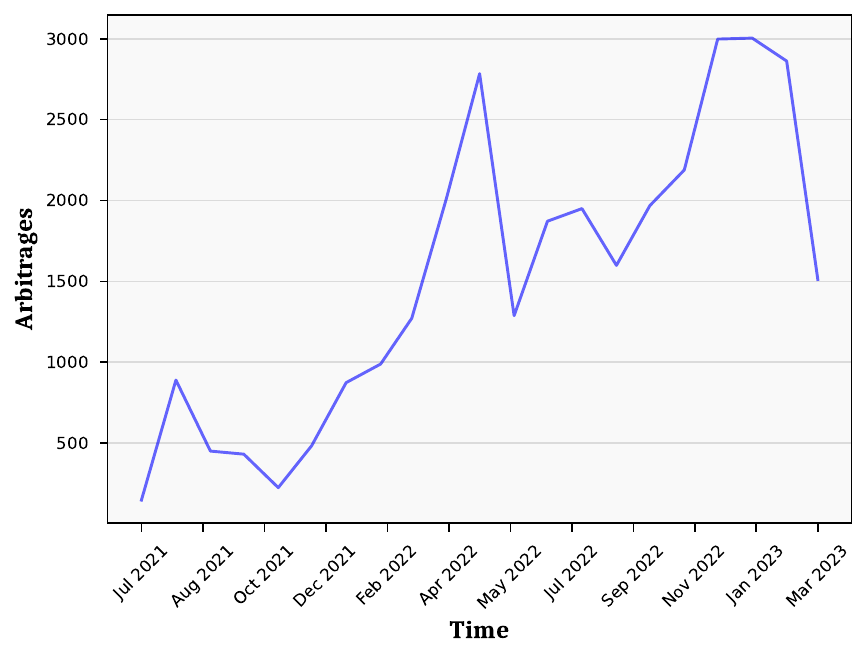}
	\vspace{-2mm}
	\caption{Arbitrages over time}
	\label{fig:arbitrage_profit_timeline}
        \vspace{-8mm}
\end{figure}

To understand the frequency of arbitrages, we plotted the number of such \tx s over time in \Fig{fig:arbitrage_profit_timeline}.
We can see a significant increase of activity in July $2021$, which is about six months after LPs such as \nftx and \nfttwenty launched (both launched in January $2021$).

\mypar{Manual verification.}
Our analysis correctly identified all publicly reported instances~\cite{flashloan-mev-nft,flashloan-bayc-arbitrage,mev-moonbirds} that we could find.
Furthermore, we randomly selected and analyzed $\arbitrageSampledReports$ \tx s identified as instant-profit-generating.
As part of our manual analysis, we looked for the following properties of these \tx s:
\textbf{(i)} A \tx touches multiple protocols, and have to be atomic, which is why they are relayed through \smart s.
Moreover, arbitrageurs tend to conceal their logic. %
Therefore, their contracts (bots) are typically closed-source.
We examined if the \tx{} $T$ in question was triggered by a closed-source contract. %
\textbf{(ii)} We cross-referenced $T$ on \etherscan, a popular \bc explorer, that interprets the details of a \tx{} in a human-readable manner.
We verified that $T$ moves an NFT both in and out of an address.
\textbf{(iii)} We investigated if $T$ touches multiple exchanges, \nftm s, and LPs.
\textbf{(iv)} To calculate the profit of the transaction, we carefully traced the flow of \ether s and \ercfungible tokens within the \tx.
Then, we verified that the sender, or the bot contract, ended up with a positive token or \ether{} balance at the end of the \tx.
In some cases, it was not immediately apparent whether the sender accumulated a positive balance, as the final profit may have been transferred to a different address $T^\prime$ associated with the sender.
For such cases, we further checked whether some amount was unconditionally transferred to $T^\prime$.
Only transactions satisfying all the above criteria were confirmed to be a correct detections.
All $\arbitrageSampledReports$ findings turned out to be true positives.

\mypar{Limitations.}
Our approach to identify exchanges uses a simple heuristic. For example, certain MEV bots interact with accounts with unusually high numbers of \tx s.
Our approach might confuse such contracts with exchanges, and, therefore, exclude them from the set  $\mathcal{M}$ of potential profit receivers, resulting in a negative profit, and false negatives in turn.

%% file: sections/loss_minimization.tex
\section{Loss Minimization}
\label{sec:loss_minimization}

Loss minimization refers to risk management techniques that aim to reduce or mitigate potential losses incurred due to events such as market fluctuations or poor investment decisions.
In traditional markets, traders set up stop-loss orders, implement hedging strategies, diversify portfolios, or continuously adjust positions to adapt to the market conditions.

\mypar{Sale price correction.}
Sometimes, opportunistic traders (sellers) are not sure about the maximum price that a buyer would be willing to pay for an NFT.
To probe the market demand, they list the NFT at a ``bait'' price and wait to see if a buyer shows any interest in the listing.
They monitor the \ethereum mempool for any pending buy \tx s from prospective buyers, which act as a form of ``endorsement'' for the price they have listed at, and encourages them to re-list the item at a higher price.
Since the real intention of the seller is not to sell the item right away, as soon as they spot a buy \tx{} (victim), they frontrun it with a cancel order \tx{} (attack) to invalidate the listing, thus, preventing potential loss from the trade.
As a result, the seller keeps the NFT, while the buyer's \tx{} reverts.

\mypar{Quantitative analysis.}
We detected $\num{\totalFrontrunningCancelBuy}$ instances of loss minimizing trades in \numberstringnum{\numNftMarketplaces} \nftm s (\Tbl{tbl:frontrunning_backrunning_loss_minimization}).
We noticed that a \cryptopunks was initially listed at $57$ ETH at \tx{} \code{0xd84...33b} \code{}.
When a buyer was about to make a purchase, the offer was immediately withdrawn.
It was re-listed in two weeks at $69.42$ ETH, and got sold on the same day.

%% file: sections/related_work.tex
\section{Related work}
\label{sec:related_work}
In this work, we study \opt in NFT markets.
Previous work on high-frequency trading in crypto markets and market manipulation are closest to our research.
In contract to our paper, these prior studies focused on markets involving fungible tokens.

\mypar{High-frequency trading.}
Arbitrage leverages the price discrepancies between \cc exchanges to make profit.
Sophisticated algorithms~\cite{McLaughlin23,Zhou21,Wang22} have been proposed to exploit such opportunities at scale in real-time.
Liquidation is the process of selling off collateralized assets at a discounted price to cover outstanding debts.
The incentives and risks~\cite{Qin212,Perez21} of participating in this system have been thoroughly studied by previous work.
While typical lending protocols are collateralized, a flashloan is a \defi lending mechanism that allows users to borrow funds without providing any collateral.
Since flashloan provides unconditional access to large funds, Qin \etal~\cite{Qin20} has shown how it can be leveraged to maximize arbitrage profits.
Previous work~\cite{Eskandari20,Daian19} has demonstrated the challenges of high-frequency trading, like arbitrage, in presence of \fr.
Sandwich attacks, which simultaneously both frontrun and backrun victim \tx s, may generate~\cite{Zhou20} a daily revenue of over several thousand US dollars just from \uniswap.
Blockchain extractable value (BEV) is an umbrella term that refers to all those different ways, such as sandwich attacks, liquidation, and arbitrage, to make illicit or harmful profits on the \bc.
A body of work~\cite{Qin21,Piet22,Bartoletti22} has studied the prevalence of such trades, their impact, and the revenue they generate in depth.

Existing work focuses on native \cc and \ercfungible token markets, leaving out the NFT market.
To the best of our knowledge, we are the first to explore the presence of similar opportunistic trades in the NFT space, which brings unique challenges, such as speculative pricing, a broader variety of different trading actions, and the non-trivial determination of  profitability.
We aim to provide a comprehensive coverage of such \opt strategies in the context of NFTs in this work.

\mypar{Manipulation of token markets.}
An NFT rug pull refers to a fraudulent practice where the creator or seller of an NFT abruptly abandons the project, leaving investors with worthless or significantly devalued tokens.
Previous work has shown the prevalence of rug pulls~\cite{Saharoy23,Huang23,Sharma23}, and proposed models for detecting them.
Analysis models~\cite{Bartoletti20,Chen19,Chen21,Kell21} have been proposed to detect Ponzi schemes, which is a fraudulent investment scheme where the operator promises high %
yields to participants by using funds from new investors.
Pump-and-dump~\cite{Kamps18,Gandal18,Xu2019} is a manipulative practice in which attackers artificially inflate the price of a particular \cc (pump), and then quickly sell off their holdings at the inflated price (dump), leaving other investors with significant losses.
Other forms of market manipulations, such as wash trading, shill bidding, and bid shielding have also been researched~\cite{Das22,Vonwachter22,Lamorgia23,Bonifazi23,Serneels23,Liu23,Wen23}.

%% file: sections/conclusion.tex
\section{Conclusion}

In this paper, we investigate the dynamics of the NFT market, and reveal the presence of sophisticated actors who employ automated, high-frequency trading strategies, including malicious or unfair practices.
While many of the trading strategies applicable to \ercfungible tokens also apply to NFTs, the unique nature of the NFT market creates distinct opportunities for the threat actors.
We identify three broad classes of \opt strategies, \viz, acquire, instant profit generation, and loss minimization.
For each class, we first qualitatively analyze the strategies, and then quantify their prevalence and financial impact.

%% file: sections/appendix_addresses.tex
\section{EOA / Contract addresses}
\label{app:eoa_contract_addresses}

\begin{table}[h]
	\footnotesize
	
	\begin{tabular}{|l|p{1.5cm}|p{4.3cm}|}
		\hline
		
		\thead{Entity} & \thead{Purpose} & \thead{Address} \\ \hline
		
		NULL address & \makecell[l]{Minting\\source} & \makecell{\code{0x0000000000000000000}\\\code{0x0000000000000000000}} \\ \hline
		
		Burn address & \makecell[l]{Burn\\address} & \makecell{\code{0x0000000000000000000}\\\code{00000000000000000dEaD}} \\ \hline
		
		\opensea v1 & Marketplace &  \makecell{\code{0x7Be8076f4EA4A4AD080} \\ \code{75C2508e481d6C946D12b}} \\ \hline
		
		\opensea v2 & Marketplace & \makecell{\code{0x7f268357A8c25526233}\\\code{16e2562D90e642bB538E5}} \\ \hline

		\opensea v3 & Marketplace & \makecell{\code{0x00000000006c3852cbE}\\ \code{f3e08E8dF289169EdE581}} \\ \hline

		\looksrare & Marketplace & \makecell{\code{0x59728544B08AB483533}\\\code{076417FbBB2fD0B17CE3a}} \\ \hline
		
		\cryptopunks & Marketplace & \makecell{\code{0xb47e3cd837dDF8e4c57}\\\code{F05d70Ab865de6e193BBB}} \\ \hline
		
		\xy & Marketplace & \makecell{\code{0x74312363e45DCaBA76c}\\\code{59ec49a7Aa8A65a67EeD3}} \\ \hline
		
		\cryptokitties & Marketplace & \makecell{\code{0xb1690C08E213a35Ed9b}\\\code{Ab7B318DE14420FB57d8C}} \\ \hline
		
		\nifty wallet & \makecell[l]{Escrow\\wallet} & \makecell{\code{0xE052113bd7D7700d623}\\\code{414a0a4585BCaE754E9d5}} \\ \hline
		
		\tubbycats & \makecell[l]{Minting\\evasion} & \makecell{\code{0xCa7cA7BcC765F77339b}\\\code{E2d648BA53ce9c8a262bD}} \\ \hline
		
		\pranksy & \makecell[l]{Maximum\\cornering} & \makecell{\code{0xD387A6E4e84a6C86bd9}\\\code{0C158C6028A58CC8Ac459}} \\ \hline
	\end{tabular}

	\caption{Addresses of important \ethereum contracts}
	\label{tbl:contract_addresses}
\end{table}

%% file: sections/appendix_transactions.tex
\section{Transactions}
\label{app:transactions}

\begin{table}[h]
	\footnotesize
	
	\begin{tabular}{|p{1.9cm}|p{6.1cm}|}
		\hline
		
		\thead{Purpose} & \thead{\emakefirstuc{\tx} hash} \\ 
	
		\hline
		
		\makecell[l]{\azuki frontrun\\ mislisting} & {\makecell{\code{0xbdba2a984877706210aef7425824d62}\\\code{d920fb531364337731318feaa1bf4db77}}} \\ \hline
		
		\makecell[l]{\azuki frontrun\\ buy} & {\makecell{\code{0x1e8ddbf7c3e1e1e37da7811e7f67ea1}\\\code{dae9843836cd9b4e9b4da9e92d906f93b}}} \\ \hline
		
		\makecell[l]{\azuki frontrun\\ buy} & {\makecell{\code{0x1e8ddbf7c3e1e1e37da7811e7f67ea1}\\\code{dae9843836cd9b4e9b4da9e92d906f93b}}} \\ \hline
		
		\makecell[l]{\adidas\\ frontrun} & {\makecell{\code{0xa7f49a4a1da9d2e4e451402a66a5346}\\\code{f1f42c4b05e850cf175aa3e7a6dcb3e59}}} \\ \hline
		
		\makecell[l]{\clonex frontrun\\ buy} & {\makecell{\code{0xfbb69ab08428e7fa650f9cd576891b2}\\\code{97a70c1fba208192d96b20b7db7d0107e}}} \\ \hline
		
		\makecell[l]{\clonex\\ sell} & {\makecell{\code{0x37fd9c9f80892ee64769240544c1f5d}\\\code{09b392a516443cd5585264afd7c62723e}}} \\ \hline
		
		\makecell[l]{\spaceshibas\\ arbitrage} & {\makecell{\code{0x304e5959e9de78765319e5c605413fc}\\\code{f24fb8a4ab294c00b0e6c35e8acff9031}}} \\ \hline
		
		\makecell[l]{\cryptopunks\\ initial listing} & {\makecell{\code{0xd84b86a7e1a2fc05261e1d87ed18af6}\\\code{362f83190f6c4da0f8569bf0eb429233b}}} \\ \hline
		
		\makecell[l]{\cryptopunks\\ offer withdrawn} & {\makecell{\code{0xf24e6c0edf287368462d7ee944d5036}\\\code{1d9ad59a132e4941d80207806fef8df0f}}} \\ \hline
		
		\makecell[l]{\cryptopunks\\ re-listing} & {\makecell{\code{0x3ee0f2a2aca70a5e03ac1ef9dfe14a1}\\\code{e0f0c4591bd7cf5a50bc708ae5320d184}}} \\ \hline
		
		\makecell[l]{\cryptopunks\\ sell} & {\makecell{\code{0x2ff97222cdf49939d690dd517706848}\\\code{c90c11b2f7734f15e9236095f7623f8e0}}} \\ \hline
		
		\makecell[l]{\mayc reward\\ collection} & {\makecell{\code{0x62955836139fa34e8de69107b69e3f8}\\\code{10373a188eb4d6d177f71d4bef7ae8f4d}}} \\ \hline
		
		\makecell[l]{\moonbirds\\ arbitrage} & {\makecell{\code{0x0659a203bd7a97d497562b14aa18f59}\\\code{46ded50be2b14c4bbaef80f5c9c42229d}}} \\ \hline
		
		\makecell[l]{\mhclub\\ minting evasion} & {\makecell{\code{0x1f838af87000ef70e2efb539454893c}\\\code{0306eac46031836e6a94ae5241a2bf4aa}}} \\ \hline
	\end{tabular}
	
	\caption{Hashes of important \ethereum \tx s}
	\label{tbl:contract_addresses}
\end{table}